\documentclass[aps,prd,floats,amsmath,amssymb,nofootinbib,
notitlepage,
longbibliography]{revtex4-1}

\usepackage[utf8]{inputenc} 
\usepackage{latexsym} 
\usepackage{amsmath}  
\usepackage{amssymb}
\usepackage{hyperref}
\usepackage{color,graphicx}
\usepackage{epsfig}
\usepackage{booktabs}
\usepackage{graphicx}
\usepackage[caption=false]{subfig}

\usepackage{xcolor}
\hypersetup{
    colorlinks,
    linkcolor={blue!40!black},
    citecolor={cyan!80!black},
    urlcolor={red!50!black}
}

\usepackage{soul}

\begin{document}

%%%%%%%%%%%%%%%%%%
%%%   MACROS   %%%
%%%%%%%%%%%%%%%%%%

\newcommand{\victor}[1]{\textcolor{blue}{{\bf Victor: #1}}}
\newcommand{\dario}[1]{\textcolor{red}{{\bf Dar\'io: #1}}}
\newcommand{\erik}[1]{\textcolor{teal}{{\bf Erik: #1}}}

%%%%%%%%%%%%%%%%%
%%%   TITLE   %%%
%%%%%%%%%%%%%%%%%

\title{Confinement of exotic matter. I. Static solutions}
\author{Víctor Jaramillo}
\author{Erik Jim\'enez-V\'azquez}
\author{Darío Núñez}

\affiliation{Instituto de Ciencias Nucleares, Universidad Nacional
Aut\'onoma de M\'exico, A.P. 70-543, M\'exico D.F. 04510, M\'exico.}

%%%%%%%%%%%%%%%%
%%%   DATE   %%%
%%%%%%%%%%%%%%%%

\date{\today}

%%%%%%%%%%%%%%%%%%%%
%%%   ABSTRACT   %%%
%%%%%%%%%%%%%%%%%%%%

\begin{abstract}
We present the {\it ${\cal E}$-boson star}: A novel configuration of a boson star with an exotic matter nucleus; the exotic matter being described by a real massive scalar field with self-interaction term and kinetic term of the opposite sign. The other scalar field is canonical, so that  the system is similar to the material component of the quintom cosmological scenario. Considering the static spherical symmetric case, we obtain cases where both fields are distributed as concentric spheres, and others with the remarkable feature that the canonical matter is pushed outwards and obtain a shell like distribution of the canonical field, with a nucleus of exotic matter at the center. We present global properties of such ${\cal E}$-boson stars and stress the differences that these configuration have with respect to the usual boson stars. In particular, we obtain cases where the compactness goes beyond the Buchdahl limit.
\end{abstract}

%%%%%%%%%%%%%%%%
%%%   PACS   %%%
%%%%%%%%%%%%%%%%


%%%%%%%%%%%%%%%%%%%%%%
%%%   MAKE TITLE   %%%
%%%%%%%%%%%%%%%%%%%%%%

\maketitle

%%%%%%%%%%%%%%%%%%%%%%%%%%%%%%%%%%%%%%%%
\section{Introduction}
\label{Sec:intro}
%%%%%%%%%%%%%%%%%%%%%%%%%%%%%%%%%%%%%%%%

The study of the dynamics of matter different to the usual gas or fluid, is an academic subject by itself. Such studies acquire more relevance once that matter is related to possible models describing the dark components of the Universe. Indeed, several models proposed for describing the dark matter, and all the ones proposed to describe the dark energy, even including the cosmological constant one, violate the strong energy condition. Some particular models, specially those concerning dark energy fail to satisfy the null energy condition (NEC) which in turn implies the violation of the weak energy condition, 
%\cite{carroll2004spacetime}
meaning that at least some time-like observer with four-velocity $u^\mu$ measures negative energy densities $\rho=T_{\mu\nu}\,u^\mu\,u^\nu$. An example of this are the phantom models \cite{Caldwell:1999ew}, whose simplest cases consist of a scalar field with a negative kinetic term in the Lagrangian. On the other hand, quintessence and scalar field dark matter models, which encompass scalar fields with a canonical kinetic term, always satisfy the NEC \cite{Westmoreland:2013zxw}, nevertheless typical cosmological solutions with this type of matter within General Relativity, lead to periods in time with negative pressures ${T^i}_j$, as measured by an static observer, in an analogy with the fluid description \cite{Ratra:1987rm}.

In the Cosmological models the matter is characterized by their equation of state\footnote{Perhaps this is not a fortunate definition, as long as it applies directly to fluids and as mentioned above, for some models of matter the concepts of density or pressure are very peculiar and do not fit with the usual ones related to the fluid. In any case, the matter is characterized in such a way and several models have a negative $w$.} , the ratio of the pressure to the density, $p/\rho = w$. For instance in quintessence, $w$ stays above the cosmological constant value $w\geq-1$, while for the phantom field $w\leq-1$. In describing the dark energy with a dynamical equation of state, cosmological data analyses interestingly suggest that the boundary $w=-1$ is crossed in the evolution of the Universe \cite{Feng:2004ad,Zhao:2017cud}. Models that can achieve this, require two scalar field (or fluid) components \cite{Feng:2004ad,Xia:2007km} and are called quintom (see Ref.~\cite{Cai:2009zp} for a review).

Self-gravitating scalar fields that violate the NEC, such as the phantom field, are the constituent material from which hypothetical objects called wormholes are constructed \cite{Bronnikov:1973fh,Matos:2005uh,Dzhunushaliev:2017syc,Carvente:2019gkd,Chew:2019lsa}. On the other hand, canonical (complex) scalar fields build up dynamically robust gravitational solitons called boson stars \cite{Kaup:1968zz}, which are regular solutions in Euclidean topology and exist in a variety of different models, reaching compactness values comparable to those of the neutron stars. 
Also solutions to the Einstein-Klein-Gordon equations with phantom fields exist in Euclidean topology \cite{Dzhunushaliev:2008bq} 
which could well be thought of as phantom boson stars. Along the lines of the latter solutions and boson stars in this work we construct objects that combine properties of both. Indeed, in this manuscript, we present solutions to a boson star model, consisting of a canonical complex scalar field and a phantom real scalar field, both minimally coupled to Einstein gravity, where the ghost matter is confined within a bounded object made of a canonical scalar field. We call these objects ${\cal E}$-boson stars. 

We are able to capture ghost matter in two different ways: with a small distribution inside a bigger soliton of canonical scalar field; and we have also found distribution where the ghost field is present and the canonical field profile tends to zero, with a maximum located away from the center, forming a shell surrounding the ghost matter. In both cases, these solutions can be thought of as boson stars with a phantom matter core that repels (via gravity) the positive density scalar field, allowing more massive and compact solutions compared to standard boson stars. 

As mentioned in several works, for instance \cite{Nunez:2015vld}, the scalar field is not a kind of fluid and care should be made when trying to understand it as a kind of fluid. It is interesting that the dynamics described in this manuscript due to the presence of both types of fields, involves in an effective manner a kind of gravitational repulsive interaction which has no analogy to the gradients of pressure of the Archimedes' principle regarding a body submerge in a liquid.

The paper is organized as follows. In Sec.~\ref{Sec:setting} we present the action, specify the metric and scalar fields ansatz, derive the static spherically symmetric field equations and define various physical quantities that will be used in the analysis. Numerical solutions of are presented in Sec.~\ref{Sec:solutions}, starting from the known boson star and ghost field soliton solutions and then constructing composite solutions, describing global properties of the configuartions, as well as concrete individual examples. Finally we present a review of our findings and give some final remarks in Sec.~\ref{Sec:Final_remarks}. We use geometrized units, in which $G=1=c$, and the convention $(-,+,+,+)$ for the metric signature.

%%%%%%%%%%%%%%%%%%%%%%%%%%%%%%%%%%%%%%%%%%%%%%%%%%
\section{Theoretical setting}
\label{Sec:setting}
%%%%%%%%%%%%%%%%%%%%%%%%%%%%%%%%%%%%%%%%%%%%%%%%%%

%%%%%%%%%%%%%%%%%%%%%%%%%%%%%%%%%%%%%%%%%%%%%%%%%%
\subsection{Action and equations of motion}
%%%%%%%%%%%%%%%%%%%%%%%%%%%%%%%%%%%%%%%%%%%%%%%%%%

We consider a canonical complex scalar field $\varphi$ and a ghost real scalar field $\chi$,
minimally coupled to Einstein gravity. The system is represented by the action,
\begin{equation}\label{eq:action}
  \mathcal{S}=\int d^4 x\sqrt{-g} \left(\frac{1}{16\pi }R+\mathcal{L}_\varphi+\mathcal{L}_\chi\right),
\end{equation}
which contains the Einstein-Hilbert action, $R$ is the Ricci scalar, $g$ the determinant of the metric, and the canonical scalar field $\varphi$ contribution with Lagrangian,
\begin{equation}
\mathcal{L}_\varphi=-\frac{1}{2}\nabla_\mu\varphi\nabla^\mu\varphi^*-\frac{1}{2}V_\varphi
\end{equation}
as well as the phantom field contribution whose Lagrangian has the following form
\begin{equation}
\mathcal{L}_\chi=\frac{1}{2}\nabla_\mu\chi\nabla^\mu\chi-\frac{1}{2}V_\chi.
\end{equation}
The functions $V_\varphi$ and $V_\chi$ denote scalar potentials given by
\begin{equation}
V_\varphi(|\varphi|)=\mu_\varphi|\varphi|^2;\quad V_\chi(\chi)=-\mu_\chi\chi^2+\lambda\chi^4 \ .
\end{equation}
$\mu_\varphi$ and $\mu_\chi$ are the mass parameters and $\lambda$ is a (positive) self-interaction parameter. When $\chi=0$ the system described by Eq.~(\ref{eq:action}) contains the necessary ingredients to construct a mini-boson star \cite{Kaup:1968zz} and on the other hand, when $\varphi=0$ the system reduce to the action used for the phantom non-singular spherical solution \cite{Dzhunushaliev:2008bq}

Variation of the action with respect to $g_{\mu\nu}$, $\varphi$ and $\chi$ lead to the Einstein equations

\begin{subequations}\label{eq:einstein}
\begin{eqnarray}
&& R_{\mu\nu}-\frac{1}{2} g_{\mu\nu} R=8\pi T_{\mu\nu} ,
\label{eq:einstein2} \\
%&& T_{\mu\nu}=g_{\mu\nu}\mathcal{L}_\varphi-2\frac{\partial \mathcal{L}_\varphi}{\partial g^{\mu\nu}}+g_{\mu\nu}\mathcal{L}_\chi-2\frac{\partial \mathcal{L}_\chi}{\partial g^{\mu\nu}}.
&& T_{\mu\nu}=g_{\mu\nu}(\mathcal{L}_\varphi+\mathcal{L}_\chi)-2\frac{\partial (\mathcal{L}_\varphi+\mathcal{L}_\varphi)}{\partial g^{\mu\nu}},
\label{eq:einstein3}
\end{eqnarray}
\end{subequations}
and the Klein-Gordon equations
\begin{subequations}\label{eq:kg}
\begin{eqnarray}
&& \nabla_\mu\nabla^\mu \varphi=\mu_\varphi^2\varphi \ ;
\label{eq:kgphi} \\
&& \nabla_\mu\nabla^\mu \chi=\mu_\chi^2\chi-2\,\lambda\chi^3 \ .
\label{Eq:kgchi}
\end{eqnarray}
\end{subequations}
We look for static spherically symmetric solutions and thus employ the following line element,
\begin{equation}\label{eq:metric}
  ds^2=-N^2\, dt^2 + \Psi^4\left(dr^2+r^2 d\Omega^2\right) \ ,
\end{equation}
with $N, \Psi$ functions of $r$ only and $0\leq r<\infty$. For the scalar fields we assume that $\chi=\chi(r)$ and for the canonical one we take the anzats: 
\begin{equation}\label{eq:ansatzphi}
\varphi_{}=\phi(r)e^{i\omega t }.
\end{equation}%

%%%%%%%%%%%%%%%%%%%%%%%%%%%%%%%%%%%%%%%%%%%%%%%%%%
\subsection{Field equations}
%%%%%%%%%%%%%%%%%%%%%%%%%%%%%%%%%%%%%%%%%%%%%%%%%%
Substitution of the metric form \eqref{eq:metric} and the ansatz of the field \eqref{eq:ansatzphi}, into Einstein Equations \eqref{eq:einstein} yield
\begin{equation}
\begin{split}
    &\Delta_3\Psi+\pi\Psi^5\left[\left(\frac{\omega\phi}{N}\right)^2+\frac{\partial\phi^2}{\Psi^4}+ \mu_\varphi^2 \phi^2  -\frac{\partial\chi^2}{\Psi^4}- \left(\mu_\chi^2-\lambda\chi^2\right) \chi^2\right]=0,\\
    &\Delta_3 N+\frac{2\partial N\partial\Psi}{\Psi}-4\pi N\Psi^4\left[2\left(\frac{\omega\phi}{N}\right)^2-\mu_\varphi^2 \phi^2 +\left(\mu_\chi^2-\lambda\chi^2\right) \chi^2 \right]=0,
\end{split}
\end{equation}

\noindent where $\partial f:=\frac{d f}{d r}$ and $\Delta_3f:=\partial^2 f+ \frac{2}{r}\partial f$.
The Klein Gordon equation for the canonical field is
\begin{equation}    
    \Delta_3\phi+\frac{\partial \phi\partial N}{N}+2\frac{\partial \phi\partial \Psi}{\Psi}-\Psi^4\left[\mu_\varphi^2-\left(\frac{\omega}{N}\right)^2\right]\phi=0 \ ,
\end{equation}
and the equation for the ghost field is
\begin{equation}    \label{eq:chi}
    \Delta_3\chi+\frac{\partial \chi\partial N}{N}+2\frac{\partial \chi\partial \Psi}{\Psi}-\Psi^4\left(\mu_\chi^2- 2\lambda\chi^2\right)\chi=0 \ ,
\end{equation}

To solve the Einstein-Klein-Gordon system of equations \eqref{eq:einstein}-\eqref{eq:chi}, we demand regularity at the origin $r=0$ and asymptotical flatness at $r\rightarrow \infty$. We apply these boundary condition to  $\phi$, $\chi$, $N$, $\Psi$:
\begin{eqnarray}\label{eq:bc}
    \partial \phi|_{r=0}=\partial \chi|_{r=0}=\partial N|_{r=0}= \partial \Psi|_{r=0}=0\ ;\\
    \phi|_{r\to\infty} = \chi|_{r\to\infty}=0,\quad N|_{r\to\infty}= \Psi|_{r\to\infty}=1\ .\label{eq:bc2}
\end{eqnarray}

Also asymptotically flat equilibrium solutions exists only if $\omega<\mu_\varphi$. Both $\omega$ and $\lambda$ are eigenvalues to be determined iteratively together with the integration of the differential equations. 

%%%%%%%%%%%%%%%%%%%%%%%%%%%%%%%%%%%%%%%%%%%%%%%%%%
\subsection{Quantities of interest}
%%%%%%%%%%%%%%%%%%%%%%%%%%%%%%%%%%%%%%%%%%%%%%%%%%

The Komar expression which gives the gravitational mass of the star is given by the following expression
\begin{equation}\label{eq:KomarM}
  M=\frac{1}{4\pi}\int_{\Sigma_t}R_{\mu\nu}n^\mu\xi^\nu dV,
\end{equation}
where $\Sigma_t$ is an hypersurface of constant $t$, $\xi=\partial/\partial_t$ is the Killing vector associated with stationarity, $n^\mu$ the future-directed unit vector normal to $\Sigma_t$. Using the ansatz \eqref{eq:metric} we arrive at the formula $M= r^2\lim_{r\to\infty}\partial N$, for the Komar mass of the present model.

We are also interested in evaluating the compactness of the obtained solutions and for this reason we use the quantity $R_{99}$, defined as the areal radius ($\Psi^2 r$) that encompasses 99 percent of the total mass of the star. We then define the compactness as
\begin{equation}
C=\frac{M}{R_{99}},
\end{equation}
which is an important quantity in describing the properties of several compact objects and will be dealt with bellow. 

The presence of light rings in horizonless compact objects (ultracompact objects) has been deeply discussed in \cite{Cunha:2017qtt, Cunha:2022gde, Cunha:2023xrt} and are a clear signal of the (in)stability of the space time under study. These results and theorems have the NEC as a requirement for the implications of the existence of such light rings on the (in)stability of the spacetime. We will see that $\cal E$-boson stars not always satisfy this energy conditions. With these facts in mind, we have searched for the presence of light-rings in the solutions that we construct, by means of the determination of the light-ring radius $r_\mathrm{lr}$ using the procedure described in \cite{Alcubierre:2021psa}, which impose a condition on the lapse-function:
\begin{equation}\label{eq:lr}
N(r_\mathrm{lr})-r_\mathrm{lr}\left(1+2r_\mathrm{lr}\left.\frac{d\ln\Psi}{dr}\right|_{r_\mathrm{lr}}\right)^{-1}\left.\frac{dN}{dr}\right|_{r_\mathrm{lr}}=0.
\end{equation}

In this way, the energy density as measured by the Eulerian static observer with four-velocity, $u^\mu$, which coincides with the vector normal to the hypersurfaces, $n^\mu$, is an important feature of the solutions and the corresponding expression for both fields are obtained directly from the projection of the stress energy tensors on such velocity:

\begin{equation}
\rho=T_{\mu\nu}\,n^\mu\,n^\nu=\rho_\varphi+\rho_\chi,\label{eq:rhoT}
\end{equation}
with
\begin{equation}
\rho_\varphi=\frac{1}{2}\left[\frac{\omega^2\phi^2}{N^2}+\frac{\partial\phi^2}{\Psi^4}+\mu_\varphi^2 \phi^2\right] \ ,
\label{eq:rhophi}
\end{equation}
\begin{equation}
\rho_\chi=-\frac{1}{2}\left[\frac{\partial\chi^2}{\Psi^4}+\left(\mu_\chi^2-\lambda\chi^2\right) \chi^2\right] \ , \label{eq:rhochi}
\end{equation}

Finally, the geometric scalars are also useful for characterizing the solutions, analyzing the curvature of the spacetime and asses its regularity. From the line element Eq.~(\ref{eq:metric}) we obtain the following expressions for the 4D-scalar of curvature:
\begin{equation}
 R=-\frac{2}{\Psi^4}\,\left(\frac{\Delta_3\,N}{N} + 4\,\frac{\Delta_3\,\Psi}{\Psi} + 2\,\frac{\partial N}{N}\,\frac{\partial\,\Psi}{\Psi}\right),
\end{equation}
for the Weyl scalar
\begin{eqnarray}
W&=&\frac{4}{3\,\Psi^8}\,\left(\frac{\partial^2\,N - \frac{\partial\,N}{r}}{N} - \frac{2}{\Psi}\left(\partial^2\,\Psi - \frac{\partial\,\Psi}{r} -3\,\frac{\left(\partial\,\Psi\right)^2}{\Psi}\right) - 4\,\frac{\partial\,N}{N}\,\frac{\partial\,\Psi}{\Psi} \right)^2,
\end{eqnarray}
and for the Kretschmann scalar:
\begin{equation} \label{eq:K}
\begin{split}
K&=-\frac{4}{\Psi^8}\,\left(\frac{\left(\partial^2 N\right)^2 - 4\,\partial^2\,N\,\partial\,N\,\partial\,\ln\Psi}{N^2}  + \frac{8}{\Psi^2}\, \left(\left(\partial^2\,\Psi\right)^2 + 2\,\partial^2\,\Psi\,\partial\,\Psi\,\left(\frac{1}{r} - \frac{\partial\,\Psi}{\Psi}\right)\right) \right. \\
 & \left. + \frac{2\,\left(\partial\,N\right)^2}{N^2}\,\left(\frac{6\,\left(\partial\,\Psi\right)^2}{\Psi^2} + \frac{4\,\partial\,\ln\Psi + 1/r}{r}\right)
  + \frac{8\,\left(\partial\,\Psi\right)^2}{\Psi^2}\,\left( \frac{3\,\left(\partial\,\Psi\right)^2}{\Psi^2} + \frac{2\,\partial\,\Psi}{r\,\Psi} + \frac{3}{r^2}\right)\right).
\end{split}
\end{equation}

As mentioned in \cite{Carvente:2019gkd}, the canonical matter generates wells in the geometry, whereas the exotic matter produces bumps, so that the plot of the geometric scalar quantities is also useful to characterize the effects on the different types of matter on the geometry.

%%%%%%%%%%%%%%%%%%%%%%%%%%%%%%%%%%%%%%%%%%%%%%%%%%
\section{Equilibrium solutions}
\label{Sec:solutions}
%%%%%%%%%%%%%%%%%%%%%%%%%%%%%%%%%%%%%%%%%%%%%%%%%%

Before reporting numerical results for the full system (\ref{eq:action}) we first discuss and review, in Sec.~\ref{Sec:bs}, some of the general properties of the cases with $\chi=0$ and then with $\varphi=0$. Next, in Sec.~\ref{Sec:qs} we take into account both canonical and exotic fields to construct a new type of configuration.

In all the configurations presented in this section, we have taken $\mu_\varphi=\mu_\chi$ and named this quantity simply $\mu$. In the Appendix we present some cases where the masses are different $\mu_\varphi \neq \mu_\chi$.

%%%%%%%%%%%%%%%%%%%%%%%%%%%%%%%%%%%%%%%%%%%%%%%%%%
\subsection{Boson stars and phantom solitons}
\label{Sec:bs}
%%%%%%%%%%%%%%%%%%%%%%%%%%%%%%%%%%%%%%%%%%%%%%%%%%

\begin{figure}
\centering
  \includegraphics[width=0.49\textwidth]{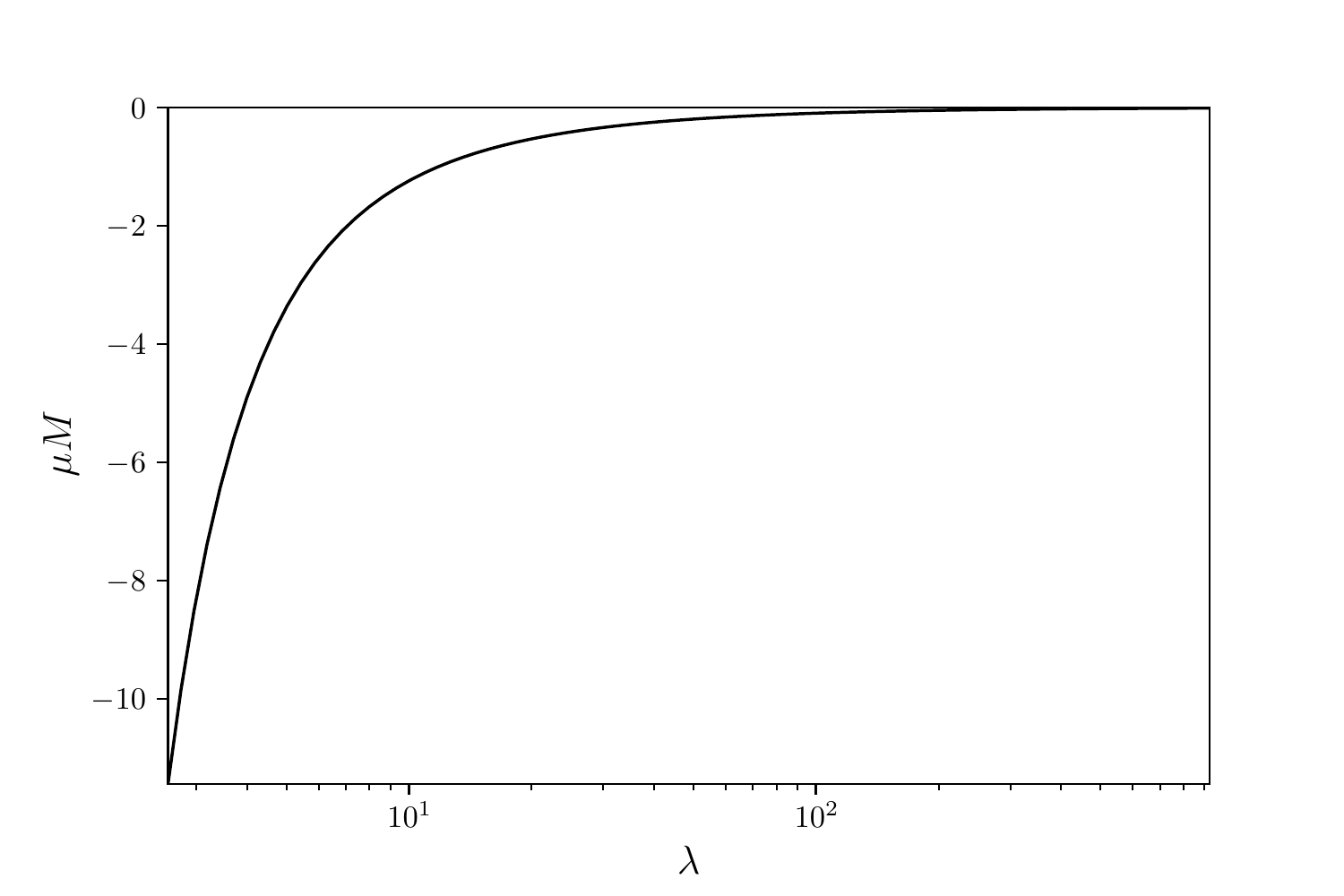}
  \caption{Non-singular solutions to
Einstein-Klein-Gordon equations with a phantom scalar field \cite{Dzhunushaliev:2008bq}. Plot of the (dimensionless) total mass of the exotic configuration, as a function of the self interaction parameter. Notice that this mass is negative for all values of $\lambda$. Such plot was originally presented in \cite{Dzhunushaliev:2008bq}.
}
\label{fig:phantom_BS} 
\end{figure}

Boson stars and the non-singular solutions with a phantom scalar field presented in \cite{Dzhunushaliev:2008bq} are very similar solutions in the sense that they both are static spherical and regular solutions of the Einstein-Klein-Gordon system in Euclidean topology. However, the latter only exists for $\lambda>0$ and have a total mass that is always negative as can be seen in Fig.~\ref{fig:phantom_BS}. Another difference is that these ghost solitons are known to be unstable.

\begin{figure}
  \includegraphics[width=0.49\textwidth]{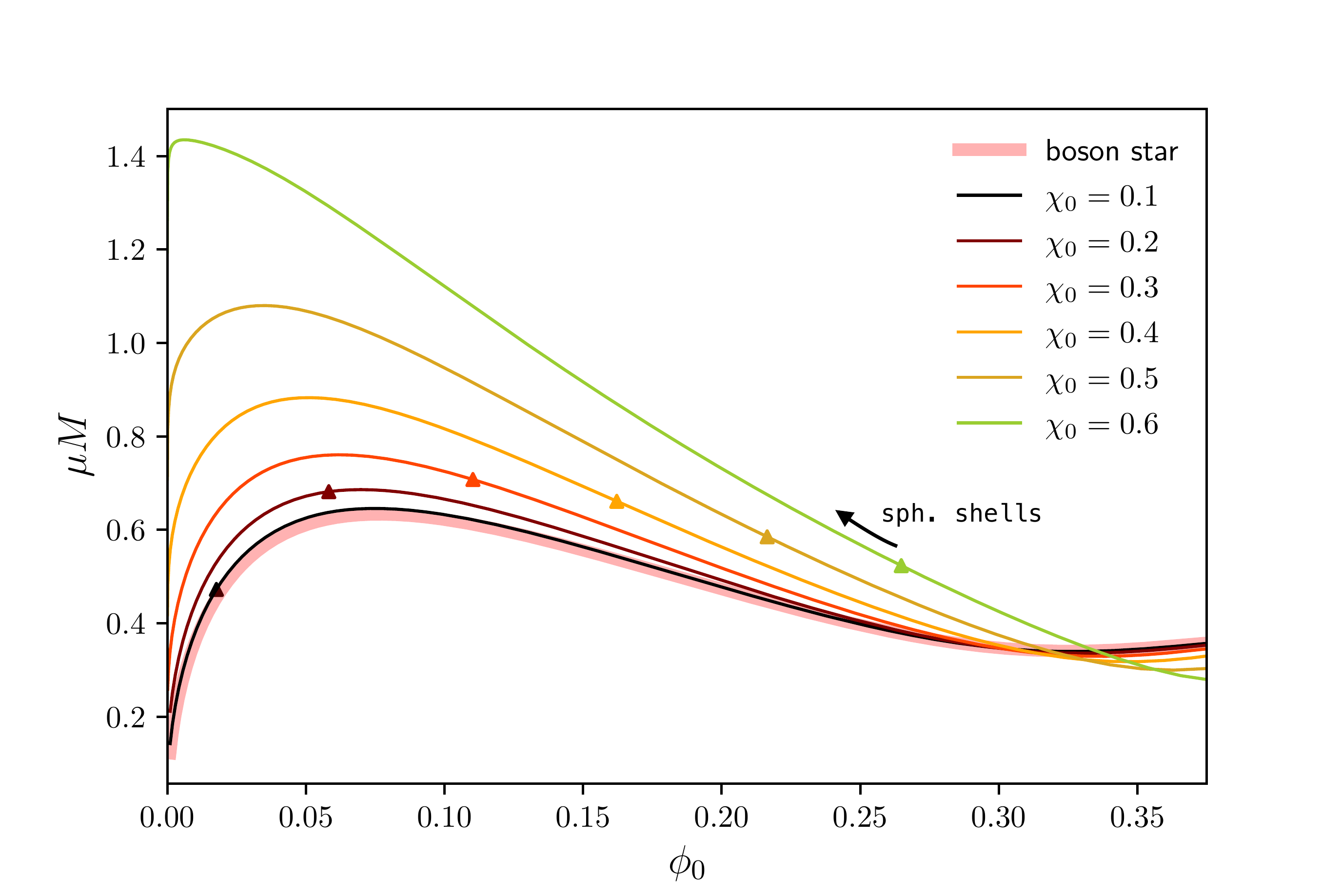} \includegraphics[width=0.49\textwidth]{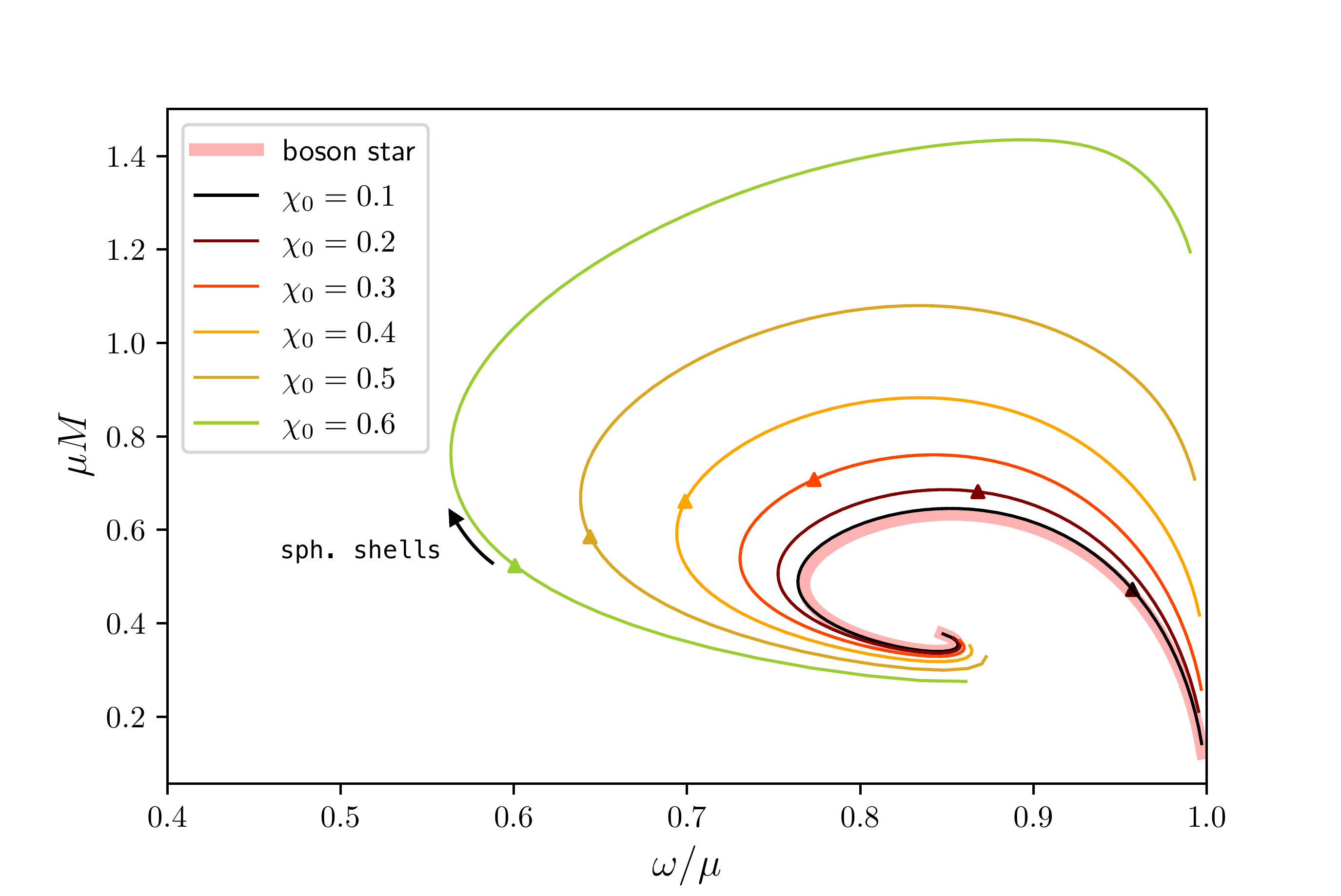}
  \caption{Left panel: Value of the canonical scalar field $\phi$ at $r=0$ \textit{vs}. total mass. Right panel: Total mass \textit{vs}. frequency. Triangles locate the transition point between $r_{\max(\phi)}=0$ (concentric spheres)
  %, left side in the upper figure, right side from the triangle in the lower one,  
  and $r_{\max(\phi)}\neq0$ (shell distribution of the canonical field). The regions that connects the triangles with the corresponding $\phi_0=0$ solutions ($\omega/\mu\to1$) correspond to the sell-like $\mathcal{E}$-boson stars.
}
\label{fig:omegavsM} 
\end{figure}

Boson stars on the other hand are stable in a region of the parameter space (see Ref.~\cite{Liebling:2012fv, Shnir:2022lba} for reviews). Given certain model, they are parameterized by the value of the scalar field at the center of the star, $\phi_0:=\phi|_{r=0}$, the family of solutions is separated by certain configuration that maximizes the value of the mass as can be seen from the thick red line at the left panel of Fig.~\ref{fig:omegavsM}. In the right panel of this same figure it can be seen how mini-boson stars, which are recovered when $\chi=0$ in \eqref{eq:action}, spiral to a limit solution when mass is plotted versus frequency.

It is worth mentioning that interesting multi-field solutions for boson stars have been reported in the literature using a complex and a real scalar field (see also \cite{Alcubierre:2018ahf, Sanchis-Gual:2021edp,Jaramillo:2022gcq} for other boson star solutions with multiple complex scalar fields). Such is the case of Boson Stars in the Friedberg-Lee-Sirlin model \cite{Kunz:2019sgn}, in which both fields are coupled (the mass of the complex field is given by its interaction with the real field) and possess a canonical kinetic term. In this FLS-model it has even been possible to construct rotating and black hole solutions with hair.  Another example are the boson stars in non-trivial topology \cite{Dzhunushaliev:2014bya}, in which a massless real field is included, giving rise to solitonic configurations with a wormhole in the core of the star. This solutions happen to inherit the Bronnikov-Ellis wormhole instability inside the star, although the authors show that in a region of parameter space the instability can be very weak.

%%%%%%%%%%%%%%%%%%%%%%%%%%%%%%%%%%%%%%%%%%%%%%%%%%
\subsection{\texorpdfstring{$\cal E$}{E}-boson stars}
\label{Sec:qs}
%%%%%%%%%%%%%%%%%%%%%%%%%%%%%%%%%%%%%%%%%%%%%%%%%%

In generating the solutions presented in this paper, we consider dimensionless quantities constructed from re-scaling with the mass of the canonical scalar field, $\mu_\varphi$. For example $r\to\mu_\varphi r$, $\omega\to\omega/\mu_\varphi$, $\mu_\chi\to\mu_\chi/\mu_\varphi$, $M\to\mu_\varphi M$, etc. so that the solutions are obtained for arbitrary $\mu_\varphi$. Then the set of four coupled non-linear ordinary differential equations for $\phi$, $\chi$, $N$ and $\Psi$ \eqref{eq:einstein}-\eqref{eq:chi} is solved numerically subject to the boundary conditions at \eqref{eq:bc} and \eqref{eq:bc2}. 

We have used a Chebyshev spectral method with the collocation approach using 24 spectral coefficients in all solutions as well as 5 radial domains with boundaries  at $\mu r=\{0.1, 2.5, 5, 25\}$. The last domain is compactified and comprises $\mu r$ from 25 to $\infty$. The resulting nonlinear algebraic system of equations is solved using a Newton-Raphson iteration.

The composite solutions ($\mathcal{E}$-boson stars), consisting of non-zero $\phi$ and $\chi$ fields, are uniquely determined by the value $\phi_0$ (defined above) and $\chi_0:=\chi|_{r=0}$, as long as we restrict ourselves to the solutions in the base state (no nodes), for all the configurations presented in the manuscript. A first solution for the composite system has been obtained by fixing small values for $\phi_0$ and $\chi_0$ and taking as initial guess a superposition of a boson star solution $\{\phi_{\mathrm{BS}},N_{\mathrm{BS}},\Psi_{\mathrm{BS}};\omega\}$ and a phantom soliton solution $\{\chi_{\mathrm{Ph}},N_{\mathrm{Ph}},\Psi_{\mathrm{Ph}};\lambda\}$ in the following form;
\begin{eqnarray}
&\phi=\phi_{\mathrm{BS}} ,\quad \chi=\chi_\mathrm{Ph}\ ;\\
& N^2 = N_{\mathrm{BS}}^2 + N_{\mathrm{Ph}}^2 -1 ,\quad \Psi^4 = \Psi_{\mathrm{BS}}^4 + \Psi_{\mathrm{Ph}}^4 -1 \ .
\end{eqnarray}

\begin{figure}
  \includegraphics[width=0.49\textwidth]{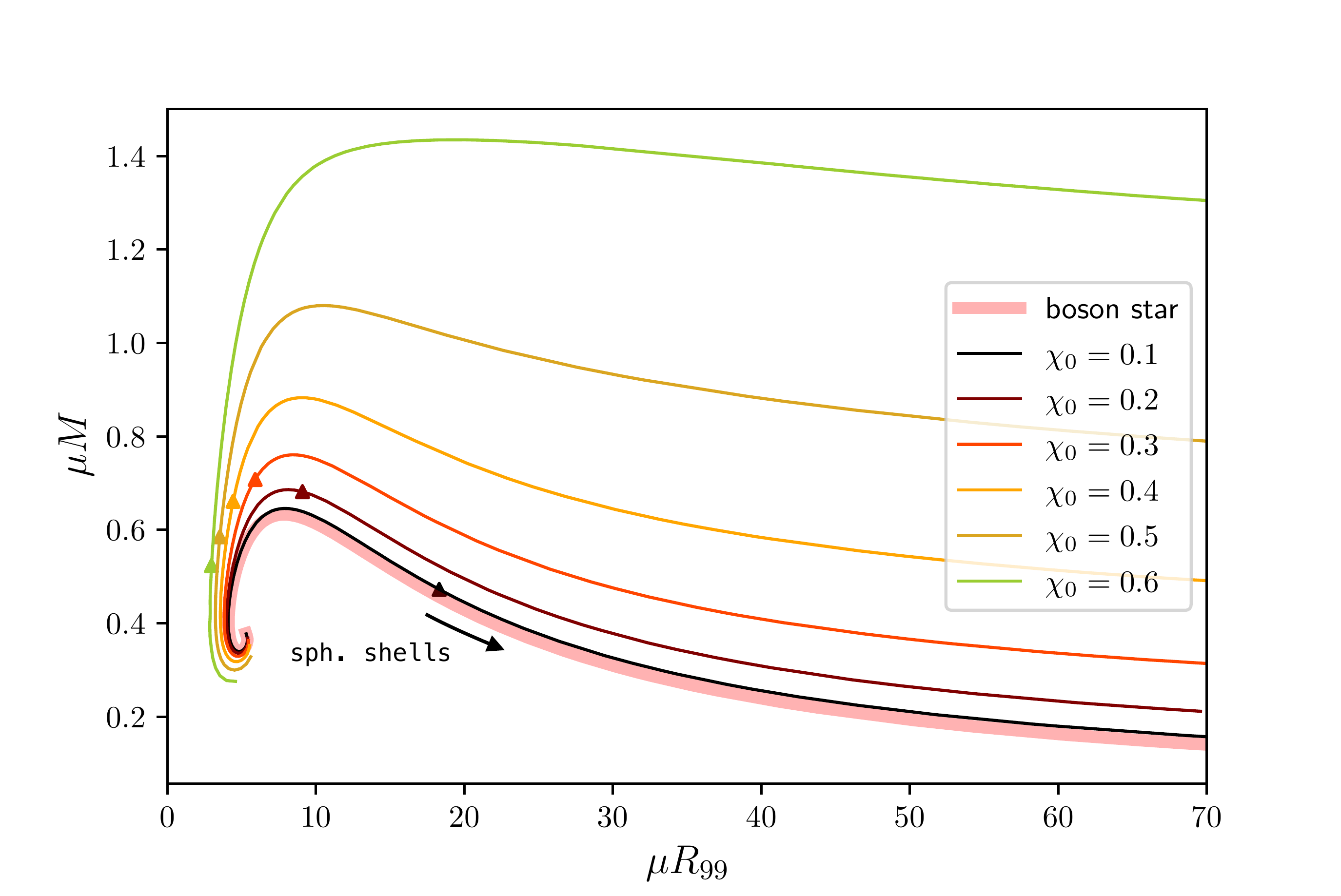} \includegraphics[width=0.49\textwidth]{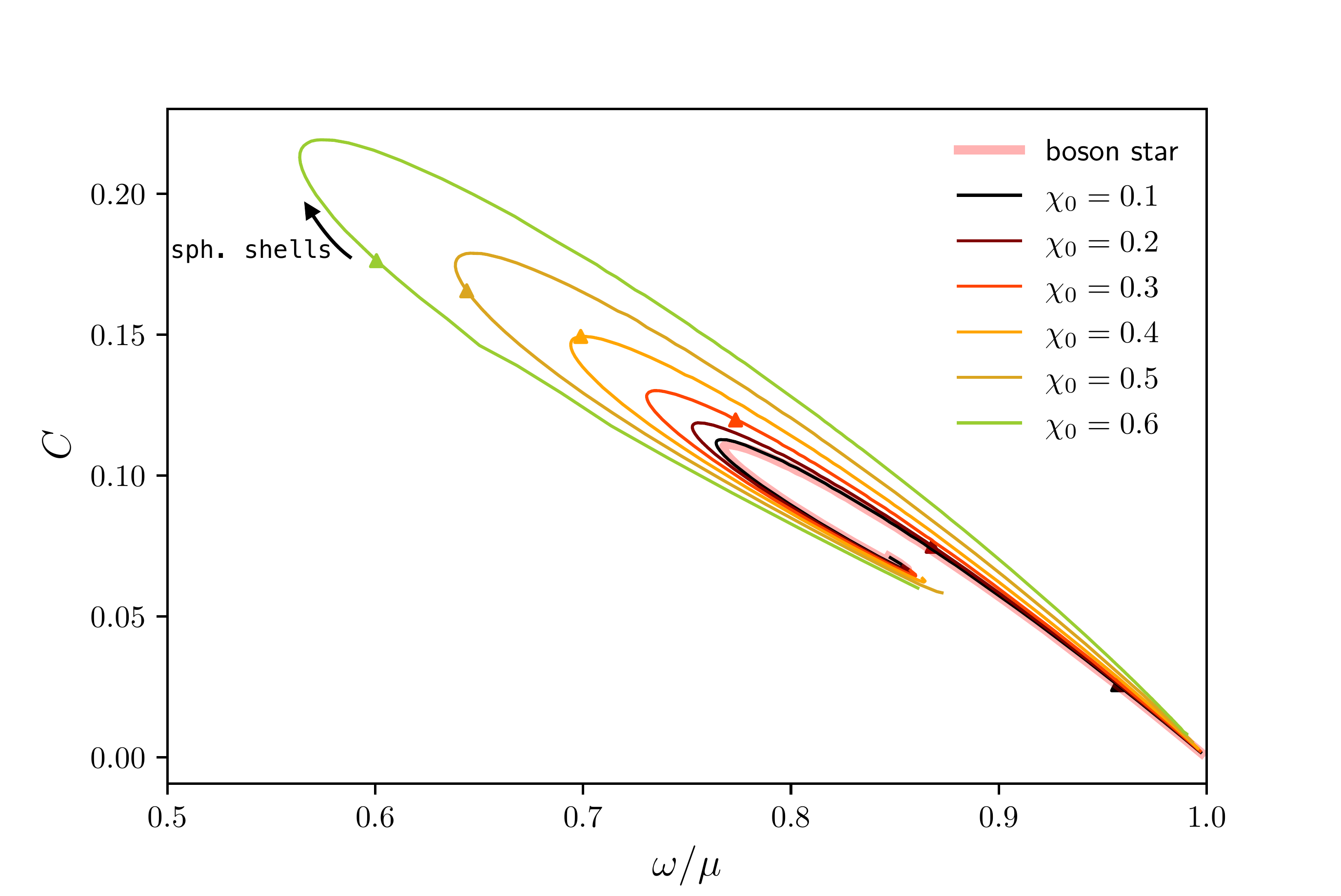}
  \caption{First panel: Radius $R_{99}$ \textit{vs}. total mass. Second panel: Frequency \textit{vs}. compactness.
}
\label{fig:RvsM} 
\end{figure}

Clearly this superposition introduce constraint violations, however the initial guess leads to configurations close to the true solution of Einstein equations, to which the code converges after a small number of steps. Then, sequence of constant $\chi_0$ solutions are obtained by slowly varying $\phi_0$. Different global quantities obtained for this families of boson star solutions are displayed in Fig.~\ref{fig:omegavsM} and \ref{fig:RvsM}, where we present the total mass of the stars vs $\phi_0$, $\omega$ and $R_{99}$. Also in the right panel of Fig.~\ref{fig:RvsM} we show the compactness as function of the frequency $\omega$. We observe that the mass and even the compactness grow notably by including a larger component of ghost matter, which, as we will see below, is stored in the core of the star.

\begin{figure}
\centering
\includegraphics[width=0.33\textwidth]{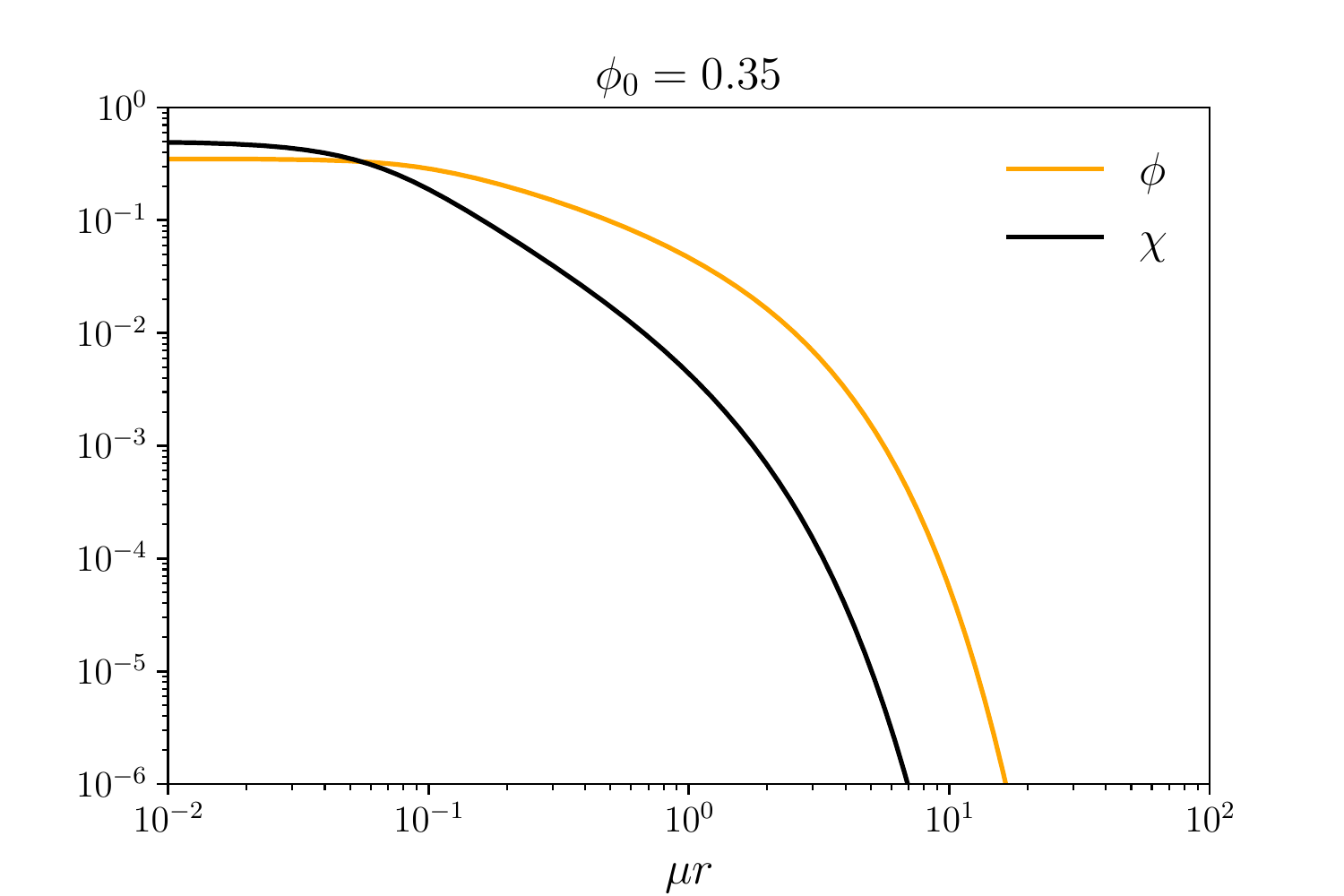}\includegraphics[width=0.33\textwidth]{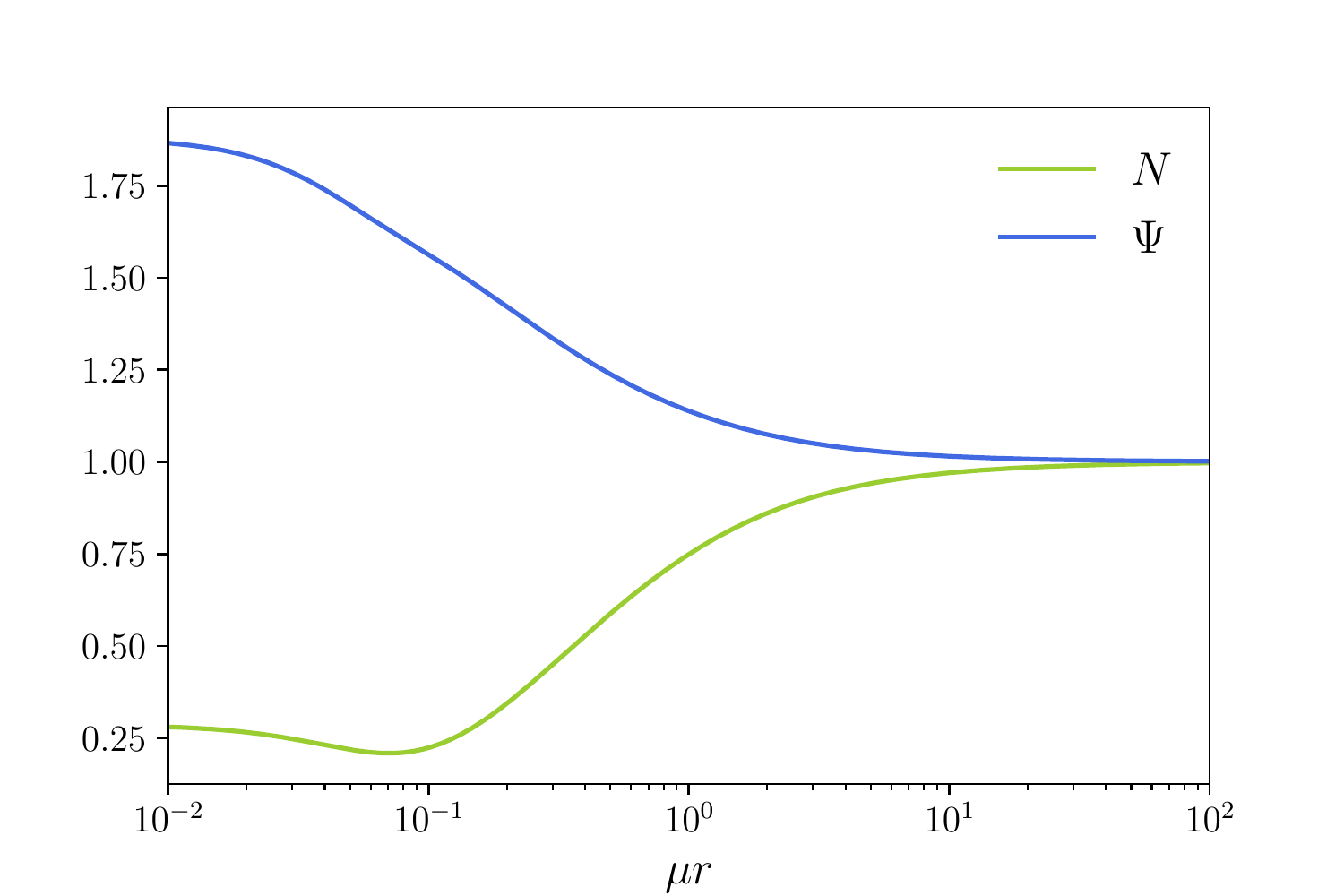}\includegraphics[width=0.33\textwidth]{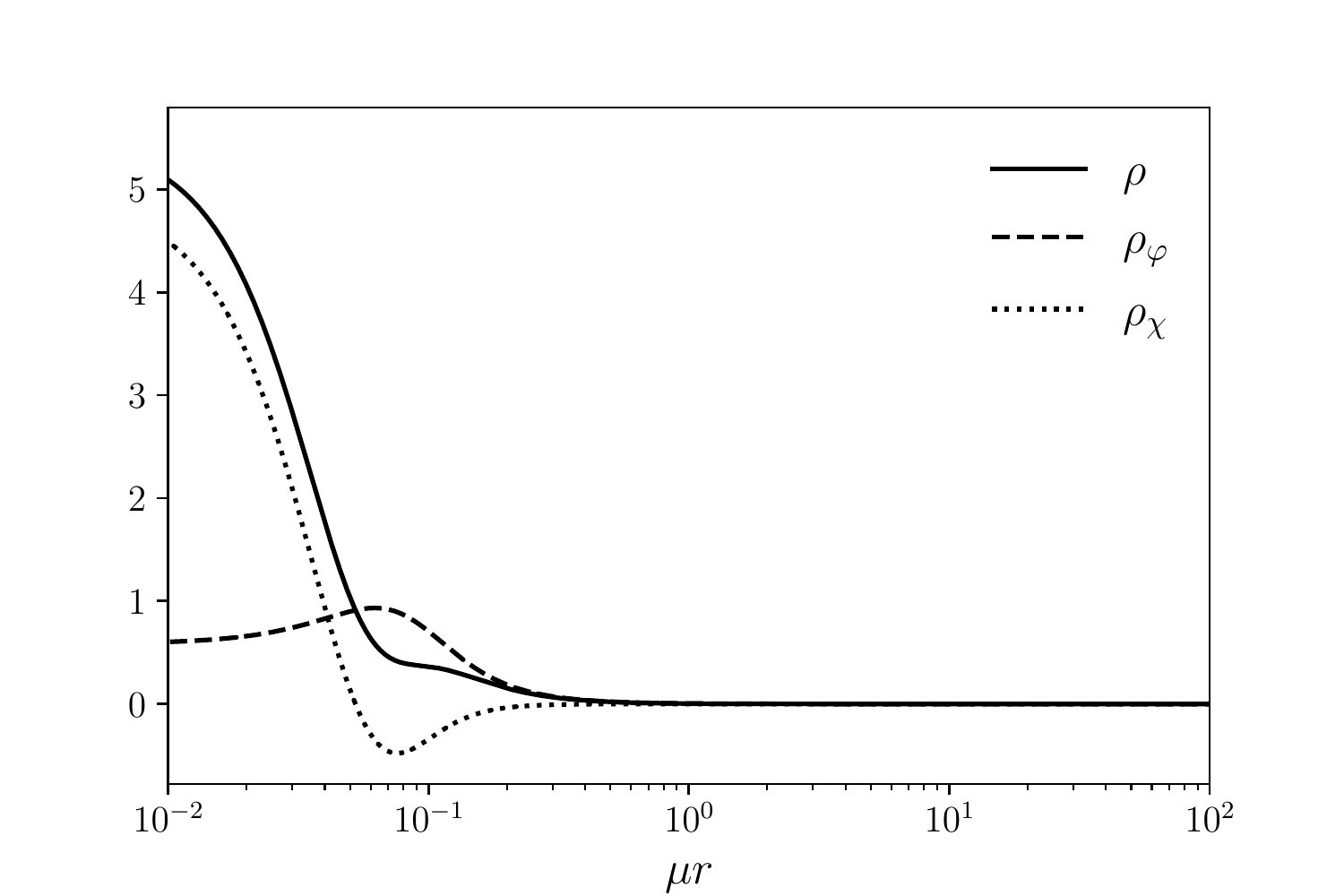}
\includegraphics[width=0.33\textwidth]{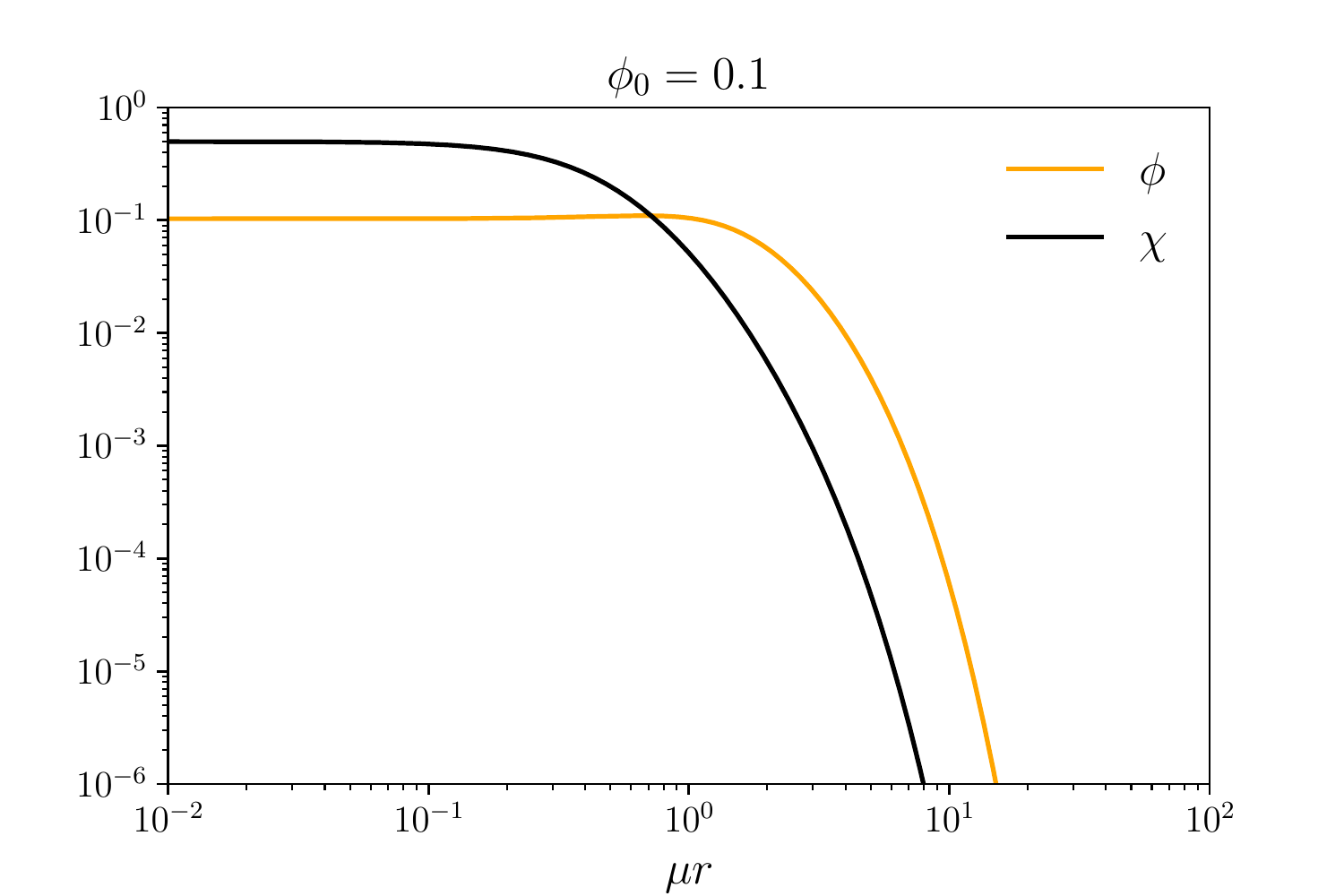}\includegraphics[width=0.33\textwidth]{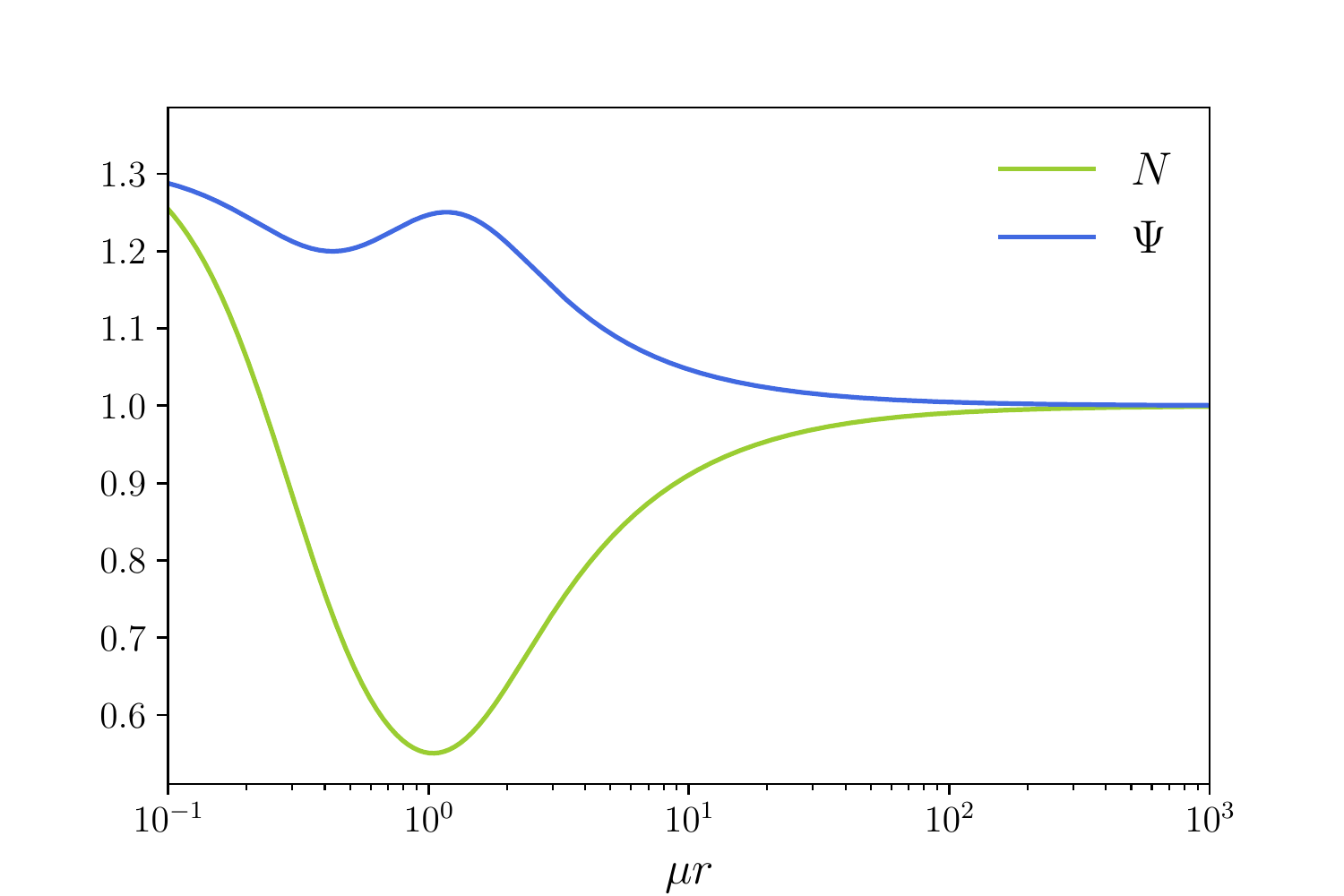}\includegraphics[width=0.33\textwidth]{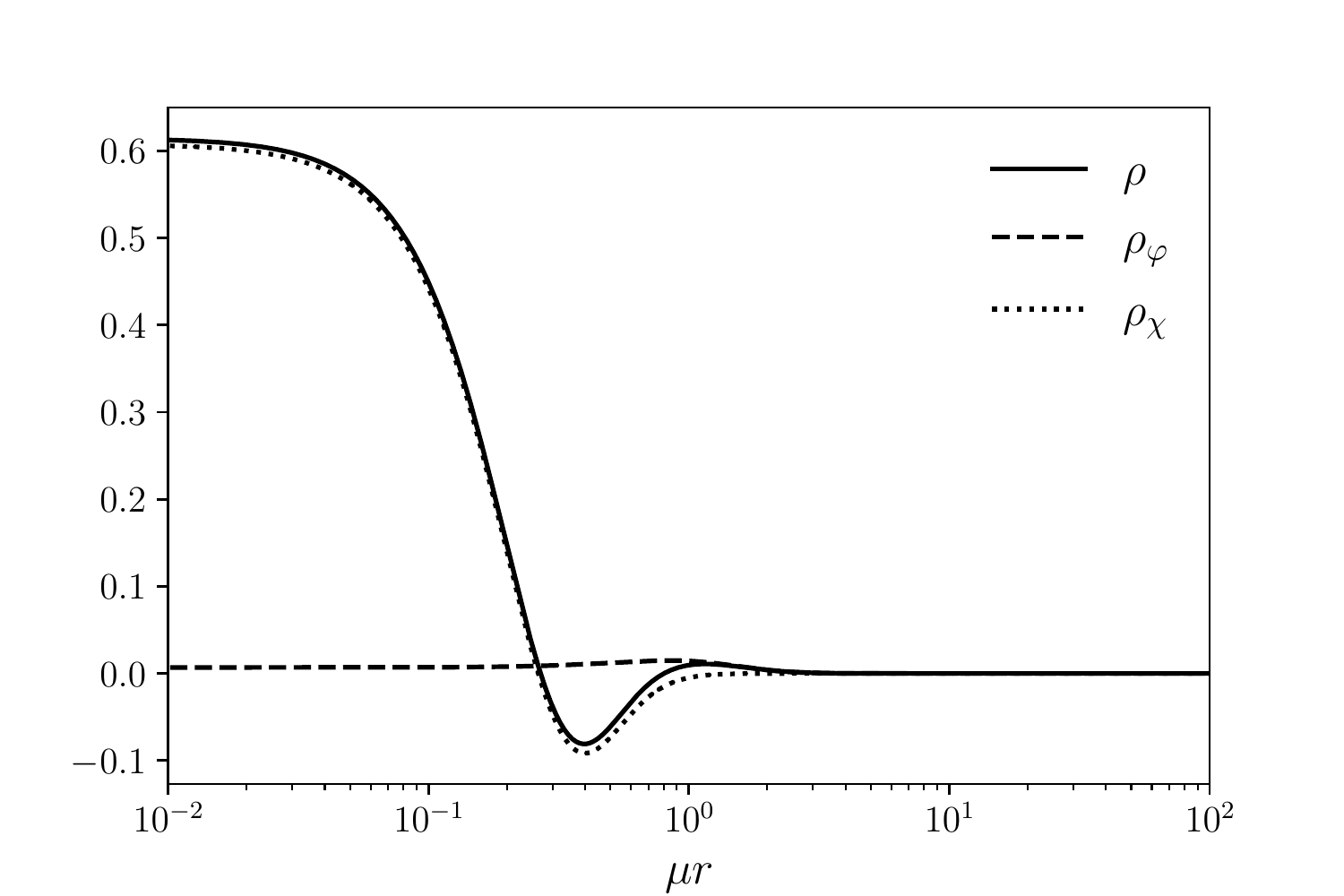}
\includegraphics[width=0.33\textwidth]{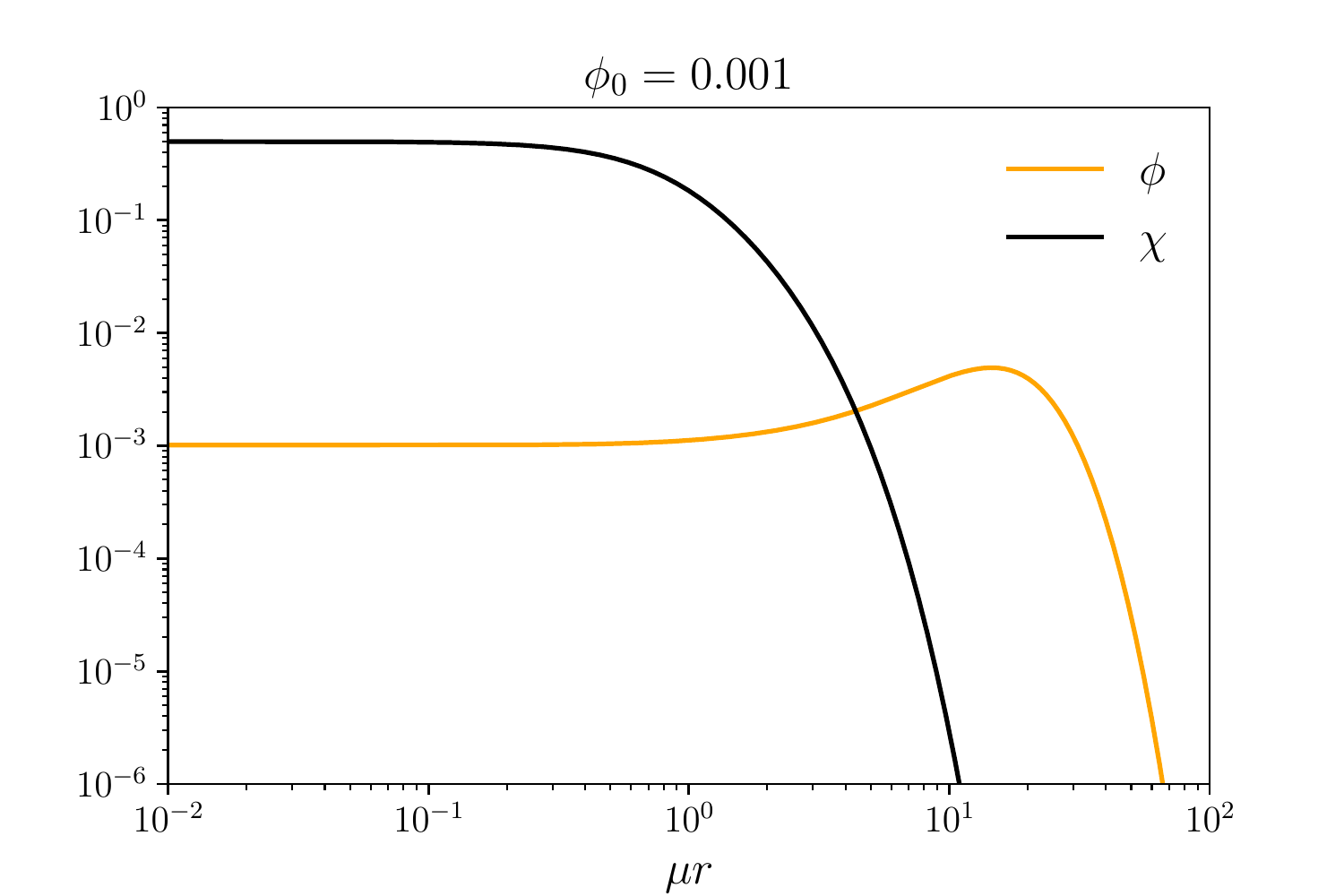}\includegraphics[width=0.33\textwidth]{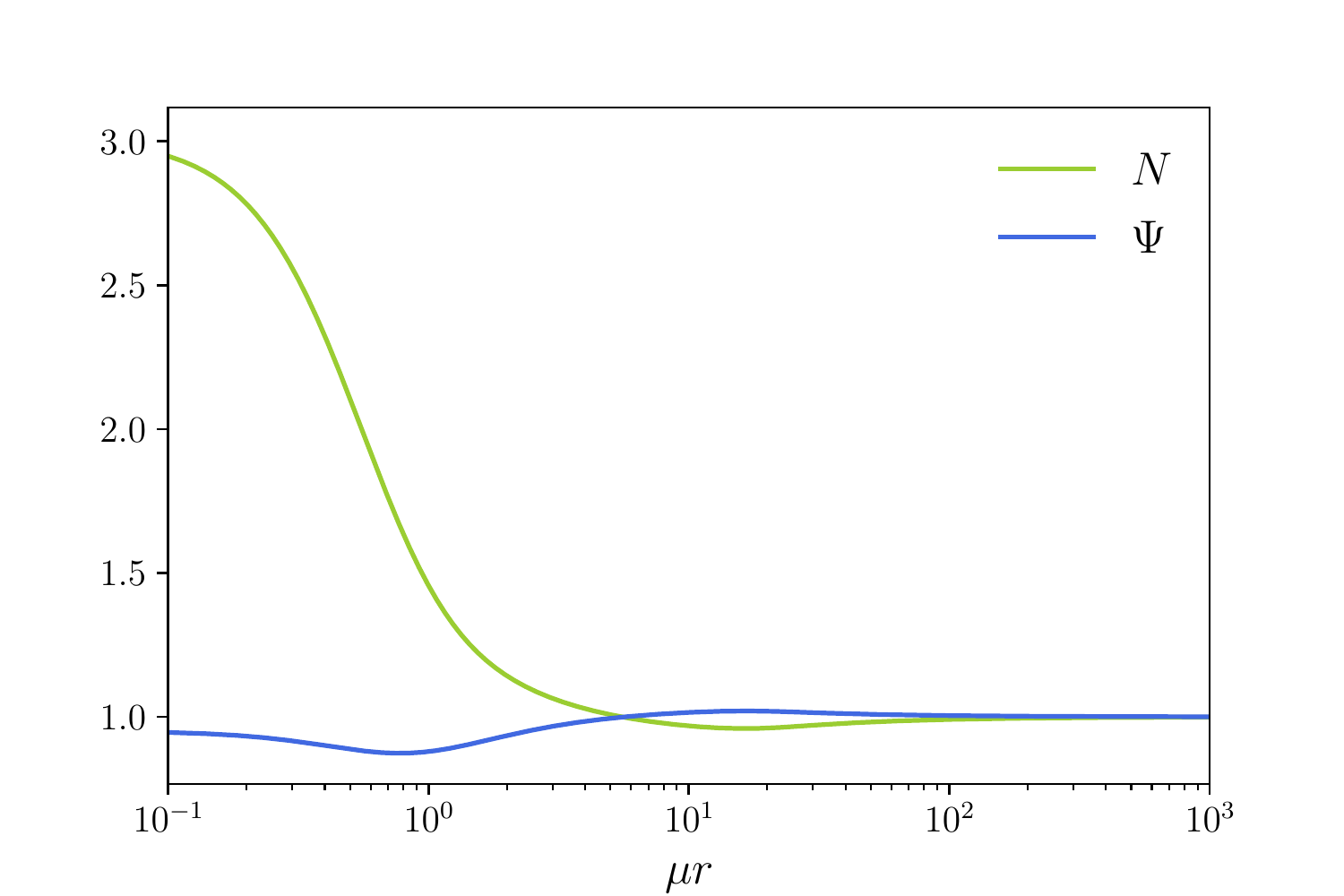}\includegraphics[width=0.33\textwidth]{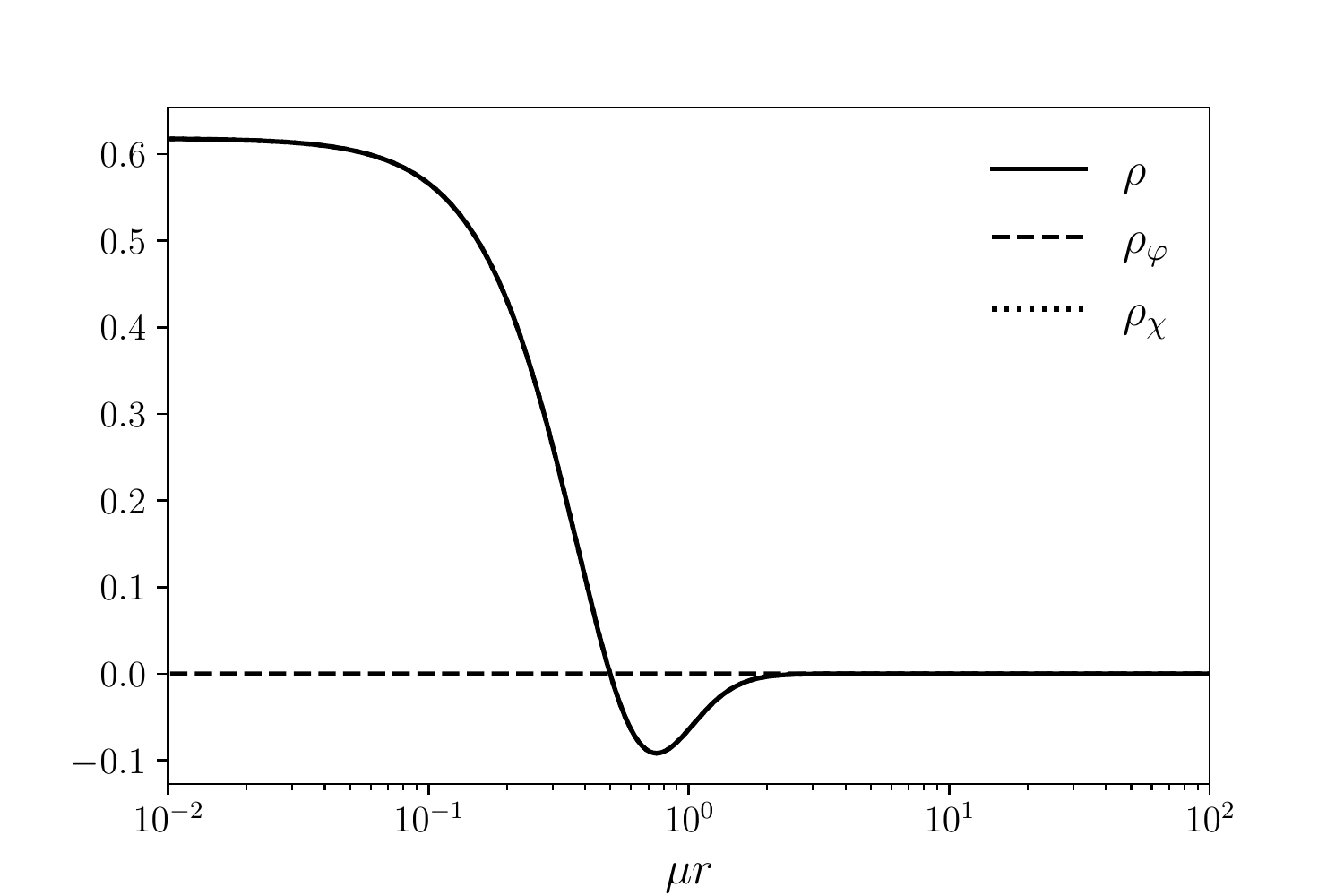}
  \caption{${\cal E}$-boson star solutions with $\chi_0=0.5$ and $\phi_0=0.35$ (upper row), 0.1 (middle row) and 0.001 (bottom row).
In the first column the fields are plotted; in the second column, the metric coefficients and in the third column the individual and total densities, all as a function of the distance to center. We notice that the energy density in the lower panel is difficult to be appreciate for the canonical field, since it is four orders of magnitude smaller.}
\label{fig:particular} 
\end{figure}

\begin{figure}
    \centering
    \hspace{0.0cm}\includegraphics[scale=0.3]{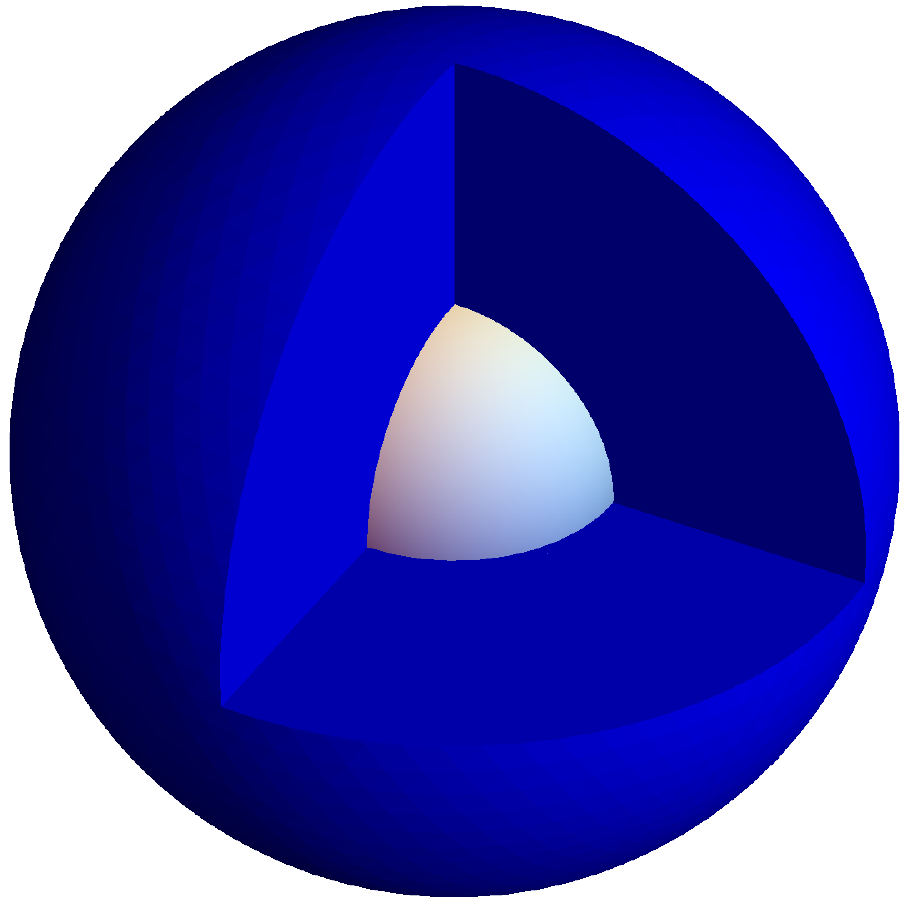}
        \hspace{3.3cm}\includegraphics[scale=0.3]{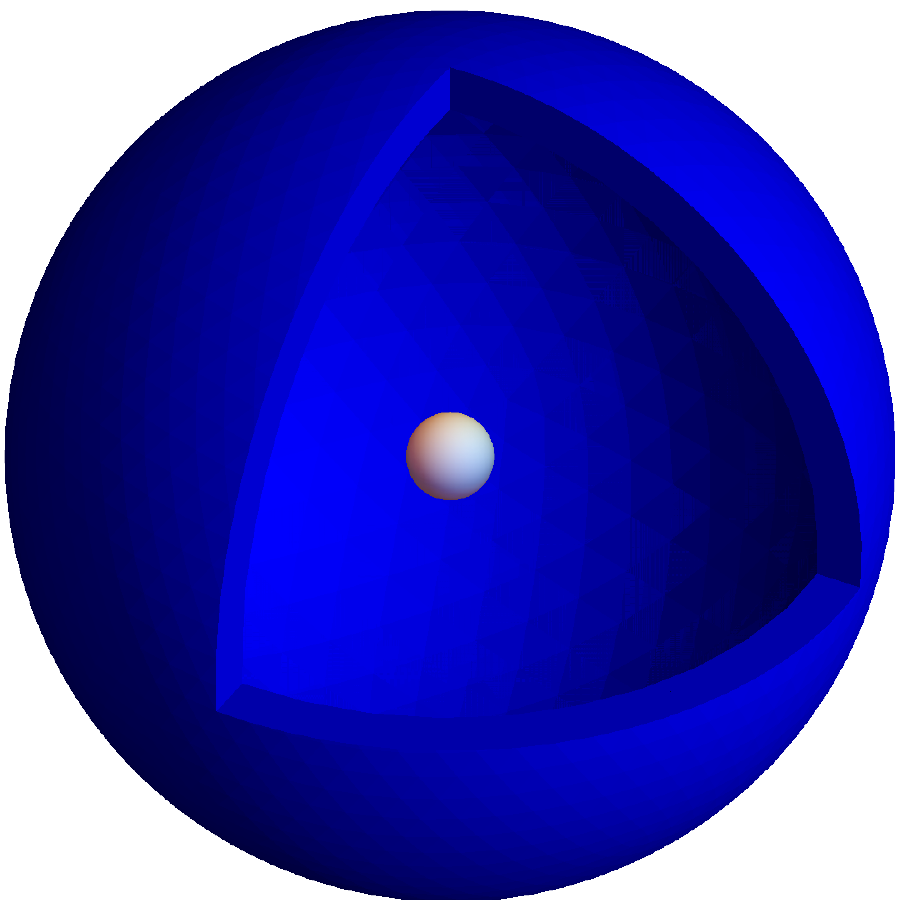}\\
    \includegraphics[scale=0.35]{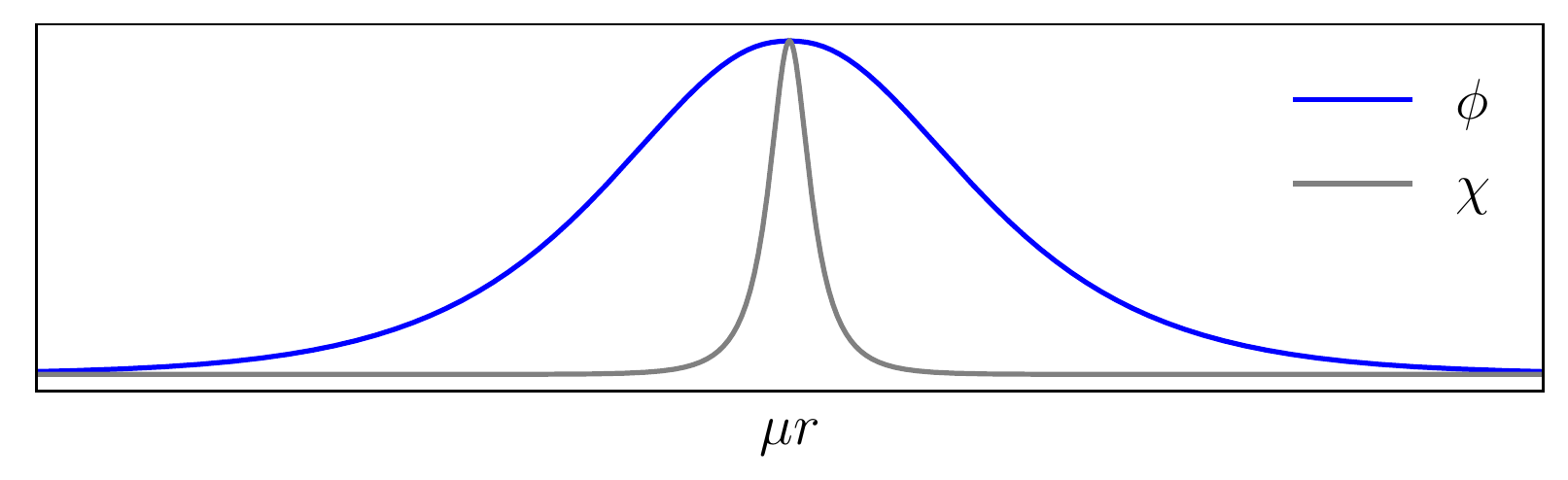}
        \quad\includegraphics[scale=0.35]{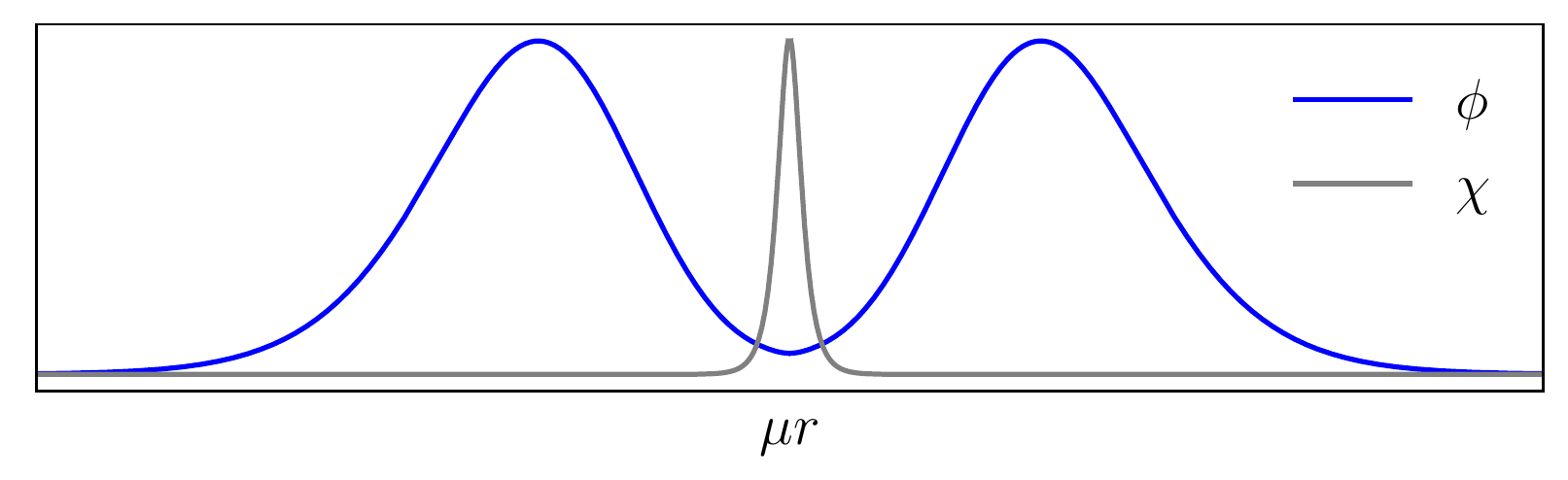}    
    \caption{Left figure: Concentric solid spheres configuration ($\chi_0=0.1$, $\phi_0=4\times 10^{-2}$). Left figure: Shell-like configuration ($\chi_0=0.6$, $\phi_0=3\times 10^{-4}$). The field profiles in the lower plots have been normalized and reflected for $-r$. Notice that the scalar fields, do not interact with one another except gravitationally and yet give rise to this type of morphologies.}
    \label{fig:Earth-like}
\end{figure}

The first and second column of Fig.~\ref{fig:particular} show solutions with $\chi_0=0.5$ and three different values for $\phi_0$. We observe an interesting feature: As the value of $\phi_0/\chi_0$ decreases (in fact, for a whole region near $\omega/\mu=1$) the $\phi$ field profile shifts its maximum away from the origin. It is also observed that for small values of $\phi_0$, the metric factors undergo qualitative changes. They go from having a monotonically increasing (decreasing) behavior for $N$ ($\Psi$) of the standard boson star, to having local maxima and/or minima. In the third column, we present the densities, the individual corresponding to each field, and the total one, Eqs.~(\ref{eq:rhochi}, \ref{eq:rhophi}, \ref{eq:rhoT}); as expected \cite{Carvente:2019gkd}, the exotic density has regions with negative values and we see that there are configuration such that the total density is positive, while others where the total density has regions with negative values; this features are also related to the distribution being concentric spheres or shells with a nucleus. Fig.~\ref{fig:Earth-like} illustrate the different morphologies of the $\mathcal{E}$-boson stars as 3D plots.

%\begin{figure}
%  \includegraphics[width=0.5\textwidth]{fig/chi-phi_1.24e-01.pdf}
%  \caption{Canonical, greater and ghost field.
%}
%\label{fig:Fatrap_1} 
%\end{figure}

%\begin{figure}
%  \includegraphics[width=0.5\textwidth]{fig/chi-phi_1.pdf}
%  \caption{Canonical, smaller and ghost field.
%}
%\label{fig:Fatrap_2} 
%\end{figure}

%\begin{figure}
%  \includegraphics[width=0.5\textwidth]{fig/chi-phi_1.00e-06.pdf}\includegraphics[width=0.5\textwidth]{fig/chi-phi_1.00e-06_phi_zoom.pdf}
%  \caption{Ghost field and zoom of the canonical
%}
%\label{fig:omegavsM} 
%\end{figure}

\begin{figure}
  \includegraphics[width=0.49\textwidth]{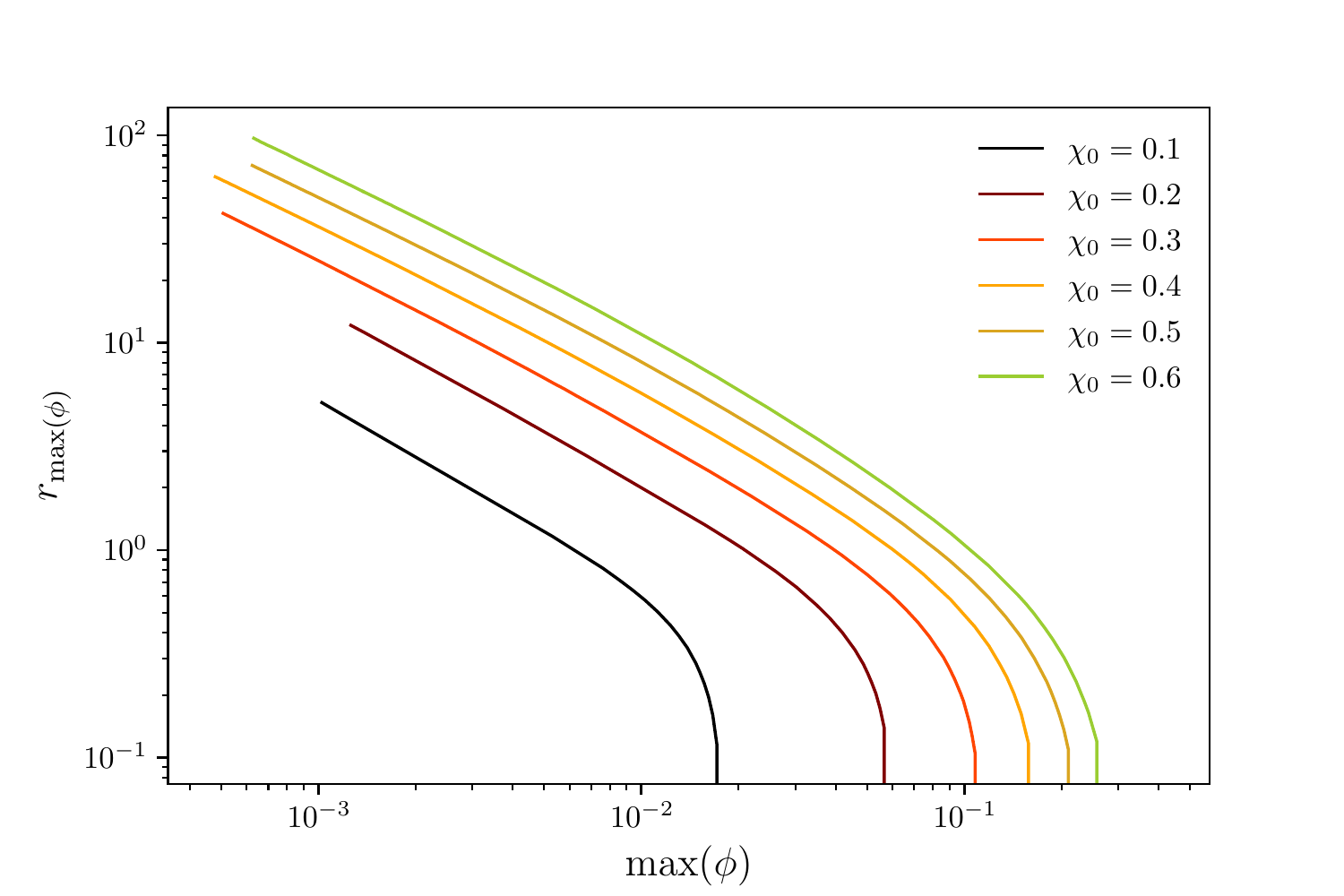} \includegraphics[width=0.49\textwidth]{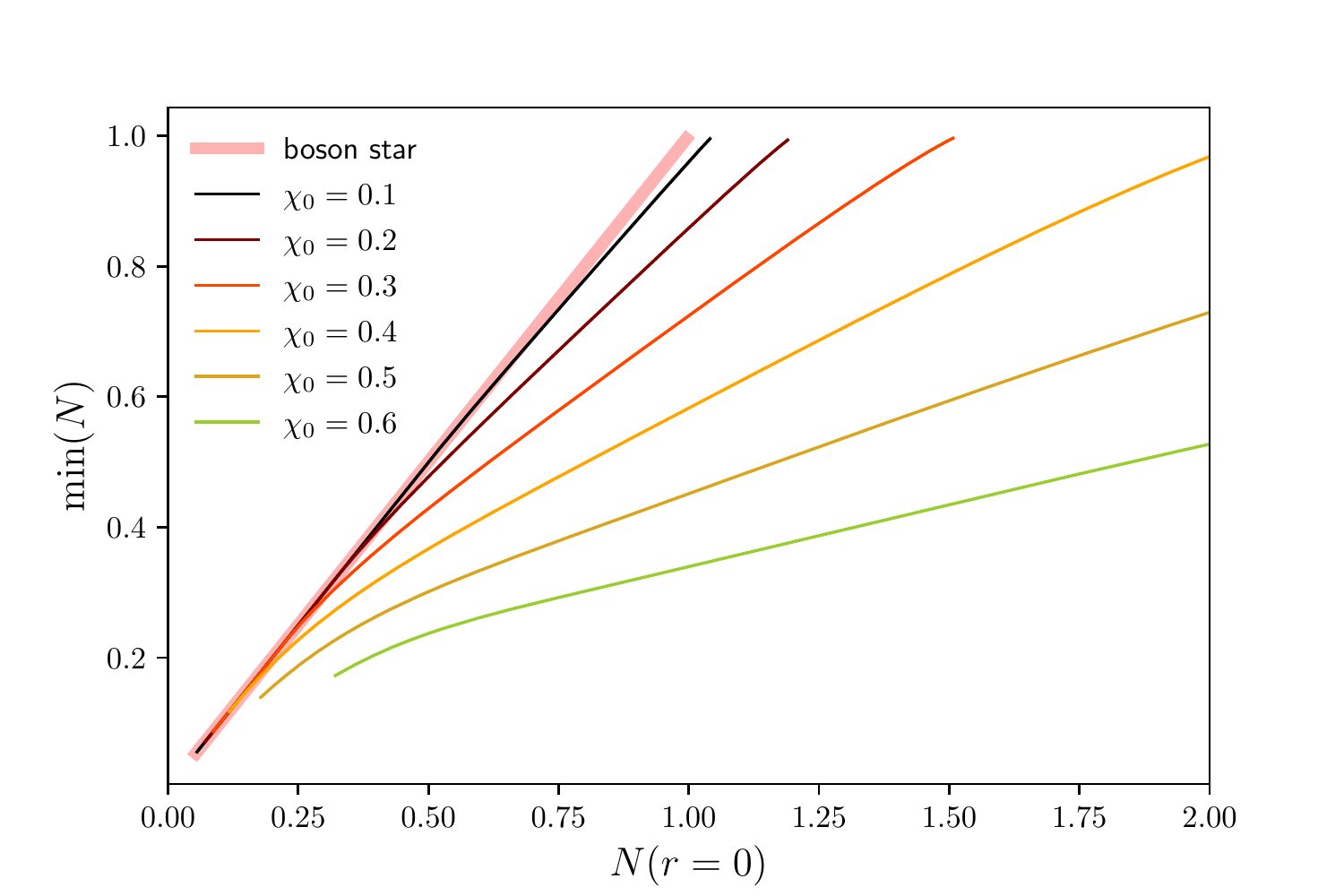}
  \caption{Left panel: Maximum of the canonical scalar field $\phi$ \textit{vs}. position at which this maximum is found. Right panel: Value of the metric function $\sqrt{-g_{tt}}$ at $r=0$ \textit{vs}. its global minimum.
}
\label{fig:rphimax}
\end{figure}

The left panel of Fig.~\ref{fig:rphimax} shows the outward shift of the maximum of $\phi$ for a set of solution families. The right panel illustrates how the minimum of $N$ ceases to be located at $r=0$, as happens in the case of the standard boson stars, $\chi_0=0$. In Figs.~\ref{fig:omegavsM}, \ref{fig:RvsM} we have indicated with a triangle the transition point from which we have solutions with the maximum of $\phi$ at $r\neq0$, \textit{i.e.}, where the canonical field changes its morphology, from a shell distribution to a solid sphere.

\begin{figure}
\centering
  \includegraphics[width=0.49\textwidth]{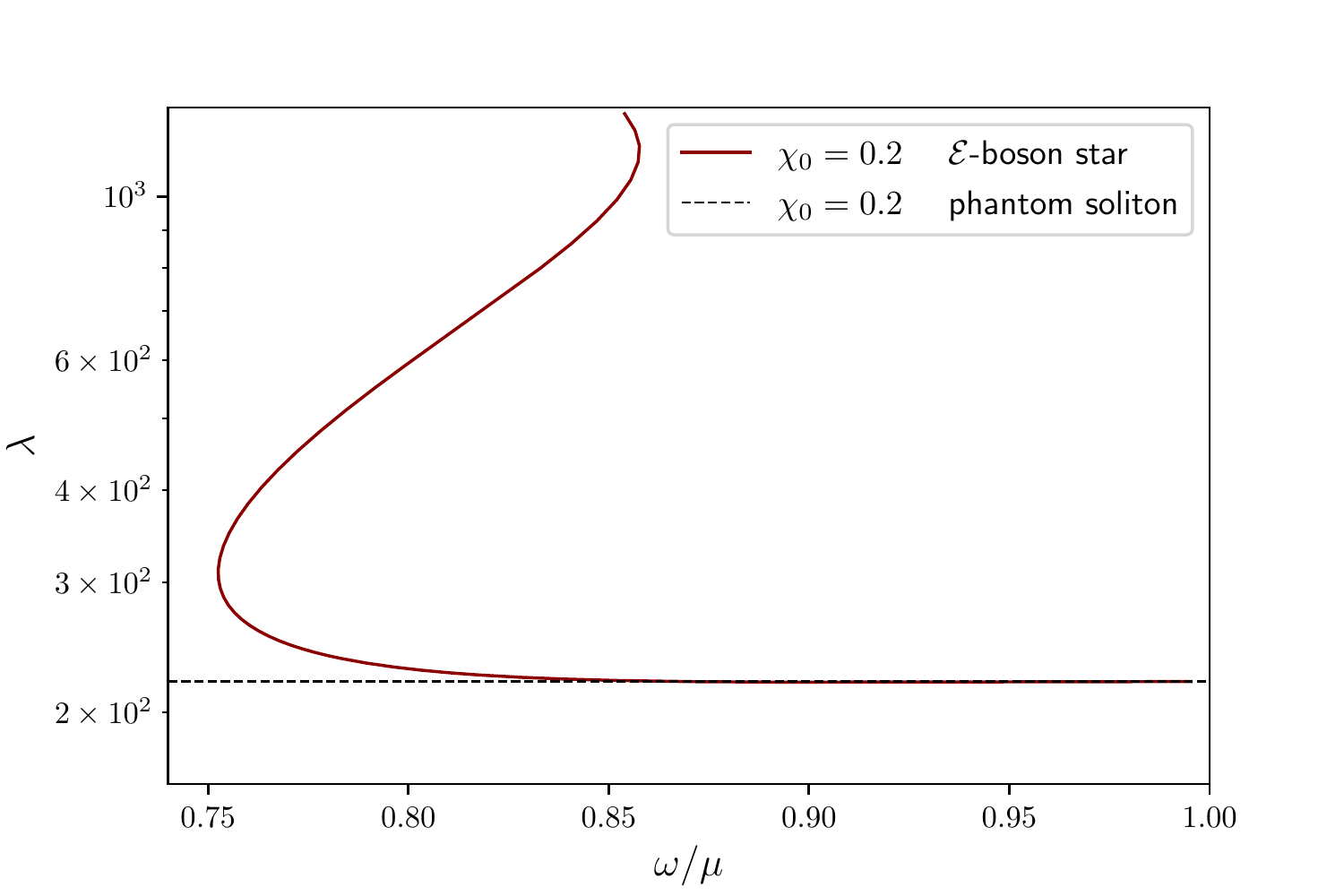}
  \caption{Eigenvalues $\omega$ and $\lambda$ for a sequence of solutions with $\chi_0$ constant.}
\label{fig:omegavslambda} 
\end{figure}

The transition to the solutions in the limit $\chi_0\to0$ (standard boson star) occurs in a simple way: the ghost field $\chi$, vanishes keeping its center at $r=0$ while the associated back-reaction of the spacetime also vanishes. However, the limit $\phi_0\to0$ is not equally straightforward as the reader can probably already anticipate from the presented results. As already mentioned, the maximum of $\phi$ moves away from the origin and the value of the absolute maximum of $\phi$ decreases (Fig.~\ref{fig:rphimax}). Also, as seen in Fig.~\ref{fig:omegavslambda}, when $\omega/\mu\to1$ the $\lambda$ eigenvalue tends asymptotically to the value that the Dzhunushaliev \textit{et al.} solution would have with the corresponding $\chi_0$. Although the canonical field energy density decreases as it approaches to this limit, the shell is of larger size and therefore contributes enough to obtain solutions whose total mass is positive. With the numerical resolution to which we have access we have not been able to find solutions with small $\phi_0\neq0$ transiting to the negative mass region, which we believe should be the case for $\omega/\mu \to 1$ in order to have a continuous transition to the phantom solitonic solution.

With respect to the energy conditions, the ${\cal E}$-boson stars have a remarkable behavior as well. Indeed,  the null energy condition states: $T_{\mu\nu}k^\mu k^\nu\geq0$, for all null vectors $k^\mu$. If the energy momentum tensor is of the form  $T_{\hat{\mu}\hat{\nu}}=\mathrm{diag}(\tau,p_1,p_2,p_3)$ with respect to an orthonormal basis (first Segr\`e type), then the NEC is satisfied if and only if $\tau+p_i\geq0$ ($i=1,2,3$) \cite{Westmoreland:2013zxw, Poisson:2009pwt}.
For the orthonormal components of the energy-momentum tensor of the present model \eqref{eq:action} and spacetime \eqref{eq:metric} we have
\begin{eqnarray}
\tau+p_1&=&\left(\frac{\omega\phi}{N}\right)^2+\left(\frac{1}{\Psi^2}\frac{d\phi}{dr}\right)^2-\left(\frac{1}{\Psi^2}\frac{d\chi}{dr}\right)^2;\label{eq:null1}\\
\tau+p_2&=&\tau+p_3\, =\, \left(\frac{\omega\phi}{N}\right)^2.
\end{eqnarray}

\begin{figure}
\centering
  \includegraphics[width=0.49\textwidth]{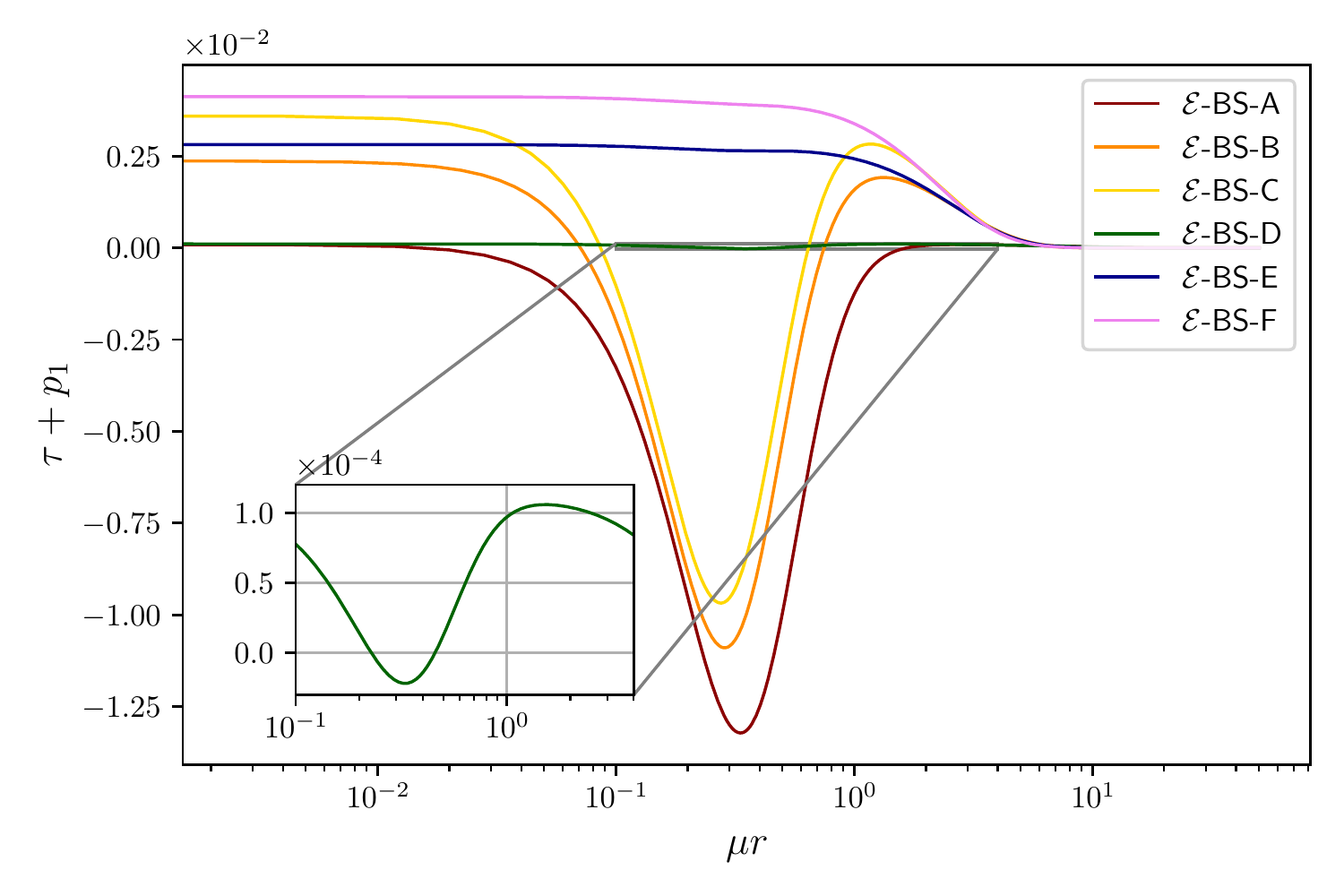}
  \caption{Null energy condition for ${\cal E}$-boson star configurations described in table \ref{tab:data}. }
\label{fig:NEC} 
\end{figure}

From these relations it can be seen that, for the Dzhunushaliev \textit{et al.} soliton \cite{Dzhunushaliev:2008bq}. when $\phi=0$, the NEC is always violated (in particular Eq.~\eqref{eq:null1}), however with respect to the ${\cal E}$-boson stars, it can be seen that there are configurations which abide the energy conditions at all points. In Table \ref{tab:data} we include data for a selection of the solutions obtained. We present two cases corresponding to boson stars, another two corresponding to the Phantom configuration and six cases of the ${\cal E}$-boson star.
Fig. \ref{fig:NEC} evaluates the NEC by plotting the quantity $\tau + p_1$ for the $\mathcal{E}$-boson star configuration in Table \ref{tab:data}. We note that among the configurations with $\chi_0=0.1$ (A, B, C) all of them violate the energy condition; the configuration A even for all r. On the other hand, for the configurations with $\chi_0 = 0.01$ (D, E, F) only D violates the NEC in a small part of the domain while E and F satisfy it at all points.

Related to the NEC violation, the high compactness obtained by the solutions (as will be seen below) and the recent results on compact objects and circular photon orbits, we wonder if the appearance of light rings happens before the critical value of the mass, in the branch that in standard boson stars corresponds to the family of stable solutions. However, what we obtain when we evaluate the condition in Eq.~\eqref{eq:lr} across fixed $\chi_0$ families of solutions, is that although the light ring "climbs" to the critical mass configuration as we increase the value of $\chi_0$, it never passes to the so-called stable branch.

\begin{figure}
\centering
  \includegraphics[width=0.49\textwidth]{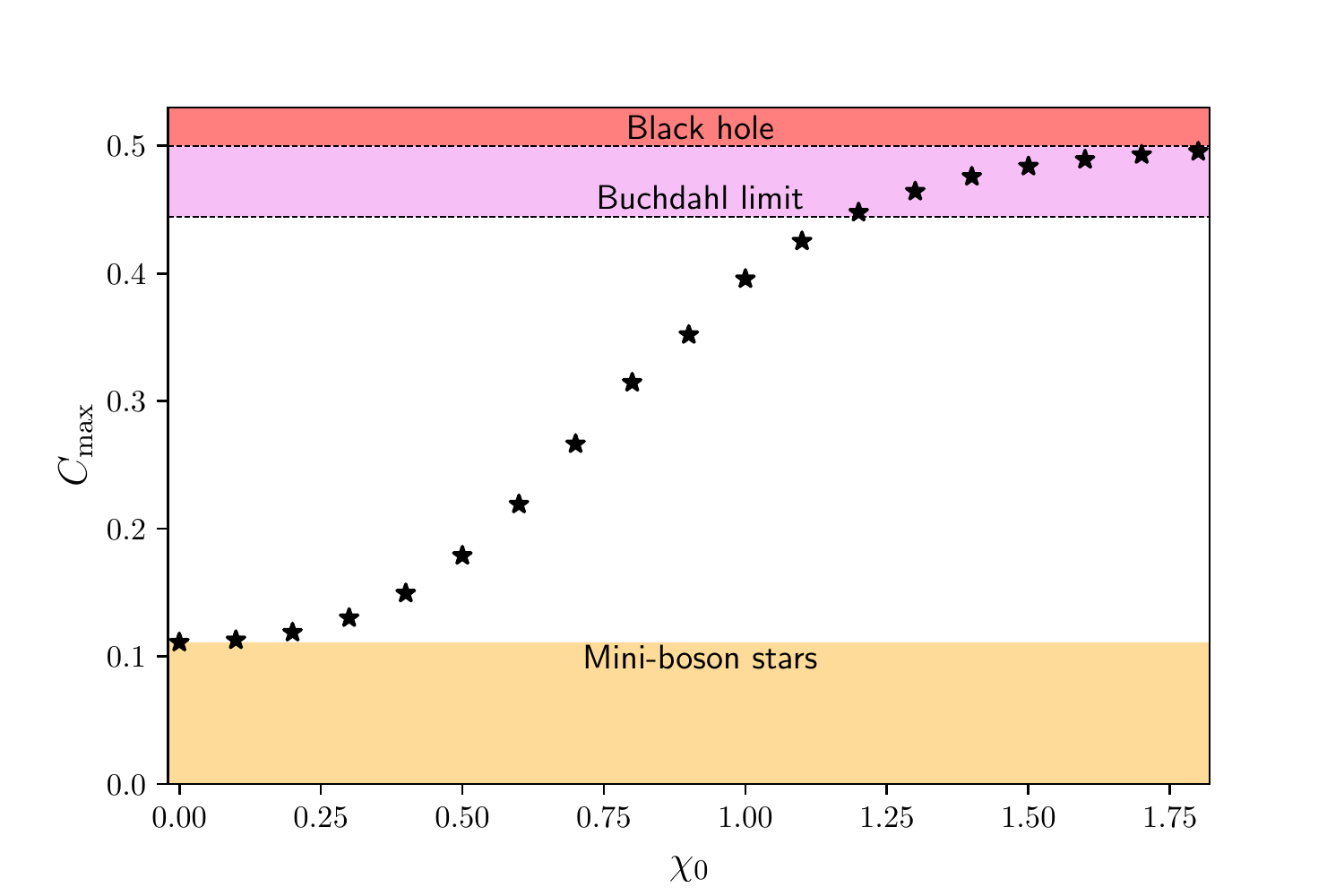}
  \caption{Maximum compactness for $\cal{E}$-boson stars with fixed values of $\chi_0$.}
\label{Fig:Cmax} 
\end{figure}

The ${\cal E}$-boson stars have also the remarkable property that the compactness of the configurations can easily rise above the Buchdahl limit as seen in Fig.~\ref{Fig:Cmax}. Indeed, the interplay of both scalars field, the exotic pushing away the canonical one, and the canonical exercising a larger pressure, provokes that the matter could be compressed to very high values\footnote{This is similar to the effect on the compactness and total mass of self-interacting boson stars \cite{Colpi:1986ye}, where the repulsive (positive) self-interaction allows to accumulate more scalar field within the star, except that in the case presented in this manuscript, the repulsion is driven by the gravitational interaction.}
As shown in Fig.~\ref{Fig:Cmax} and in the Appendix, increasing values of $\chi_0$ or $\mu_\chi/\mu_\varphi$ points to the fact that the black hole limit could be reached. There was already experience with the  $\ell$-boson stars, \cite{Alcubierre:2021psa} where the maximum compactness tends to the Buchdahl limit as the parameter $\ell$ grows, but they never rose above such limit. The $\mathcal{C}$-stars, presented in \cite{Raposo:2018rjn} are solutions which describe anisotropic fluid stars with compactness above the Buchdahl limit, due to anisotropic pressures. Gravastars are also an example of compact object solutions within general relativity with compactness above the Buchdahl limit \cite{Pani:2015tga}. See \cite{Andreasson:2007ck} for a discussion on the conditions needed to rise above such limit, and it is presented the case for a collection of non-interactive particles, the Vlasov gas, where the Buchdahl limit is not surpassed. See Ref.~\cite{Alho:2022bki} for interesting discussions on the maximum compactness of horizonless compact objects and \cite{Cardoso:2021ehg} for case of bosonic fields. 

\begin{figure}
\centering
  \includegraphics[width=0.49\textwidth]{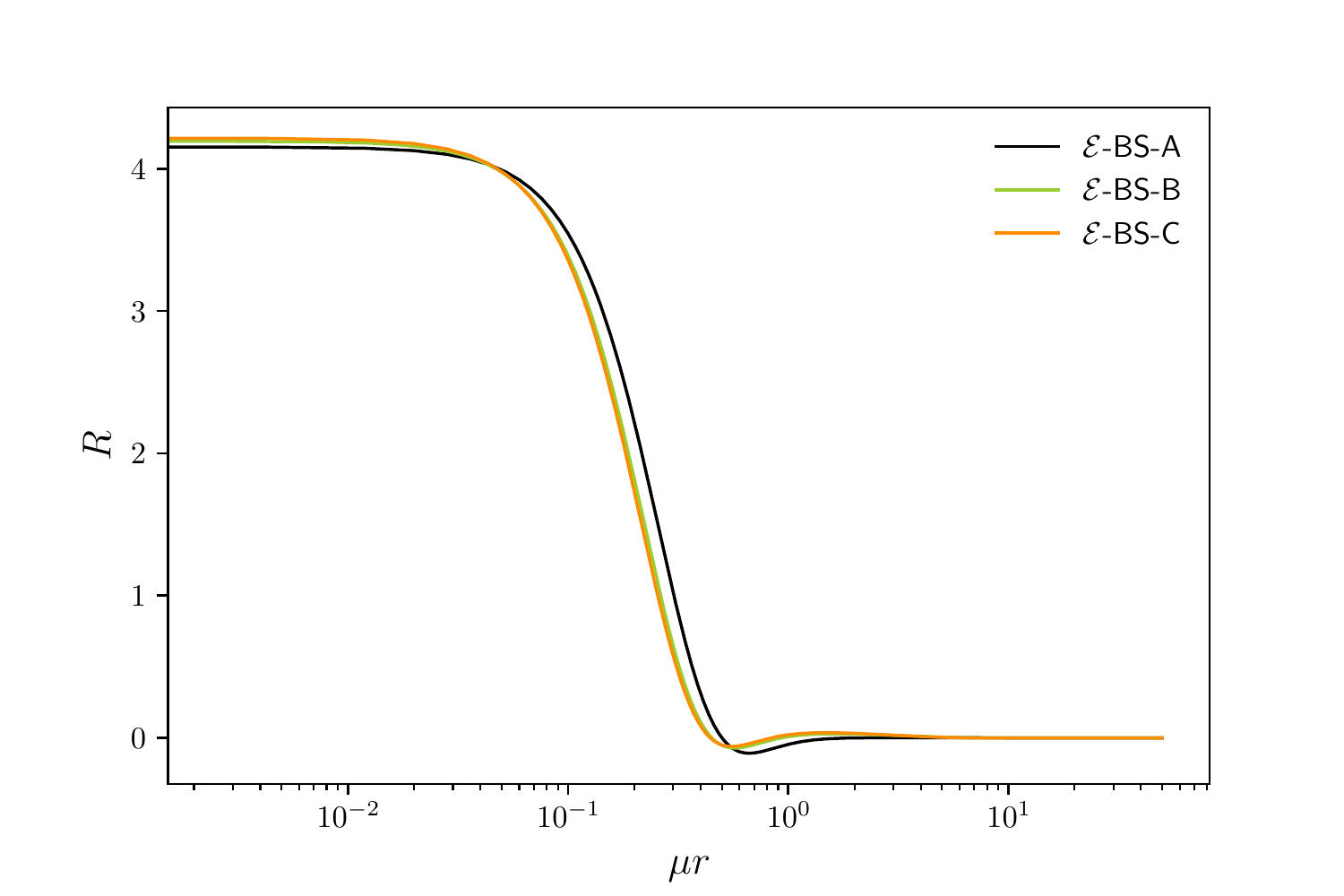}
  \includegraphics[width=0.49\textwidth]{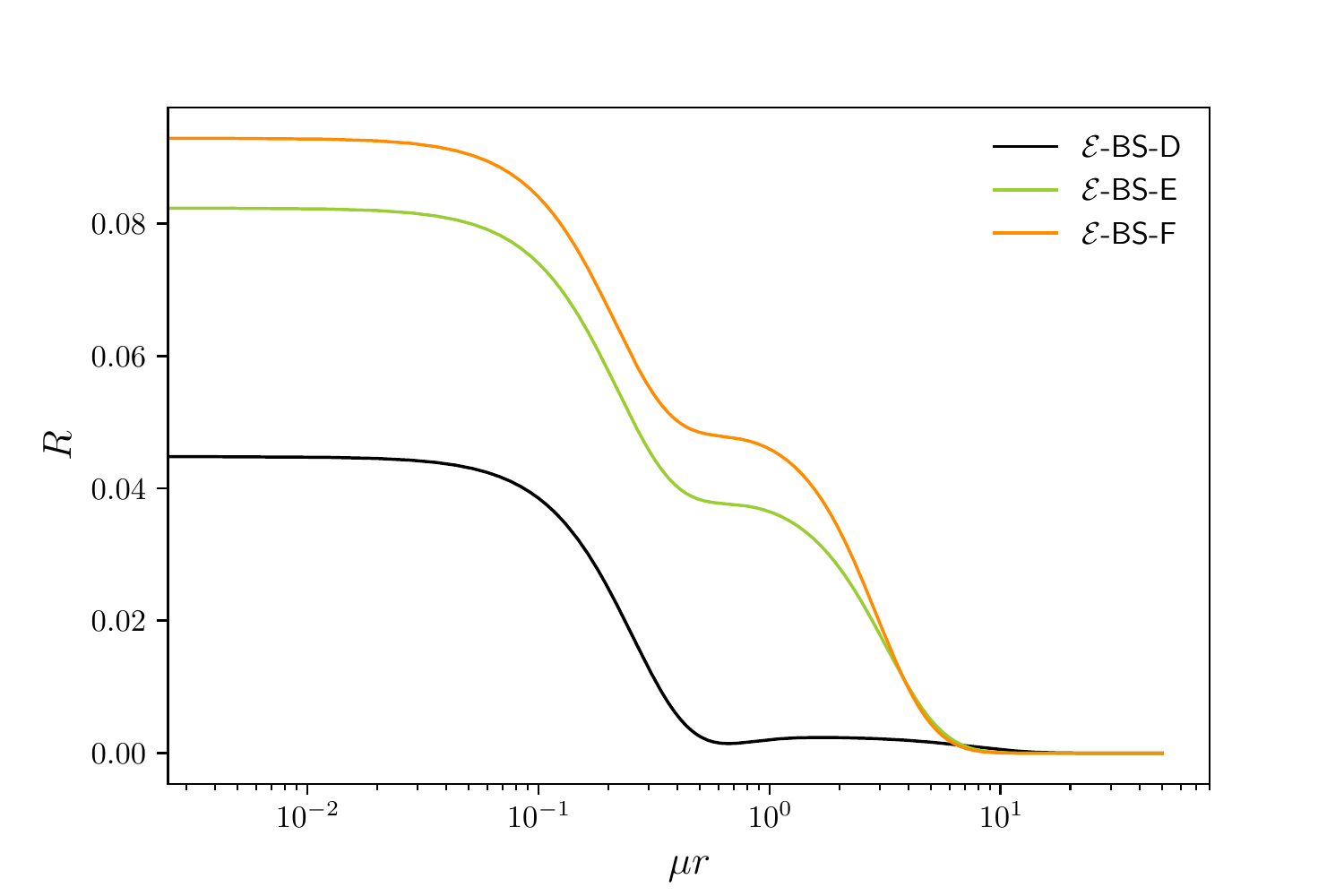}
  \caption{4D curvature scalar for ${\cal E}$-boson star configurations described in Table \ref{tab:data}.}
\label{fig:Ricci} 
\end{figure}

\begin{figure}
\centering
  \includegraphics[width=0.49\textwidth]{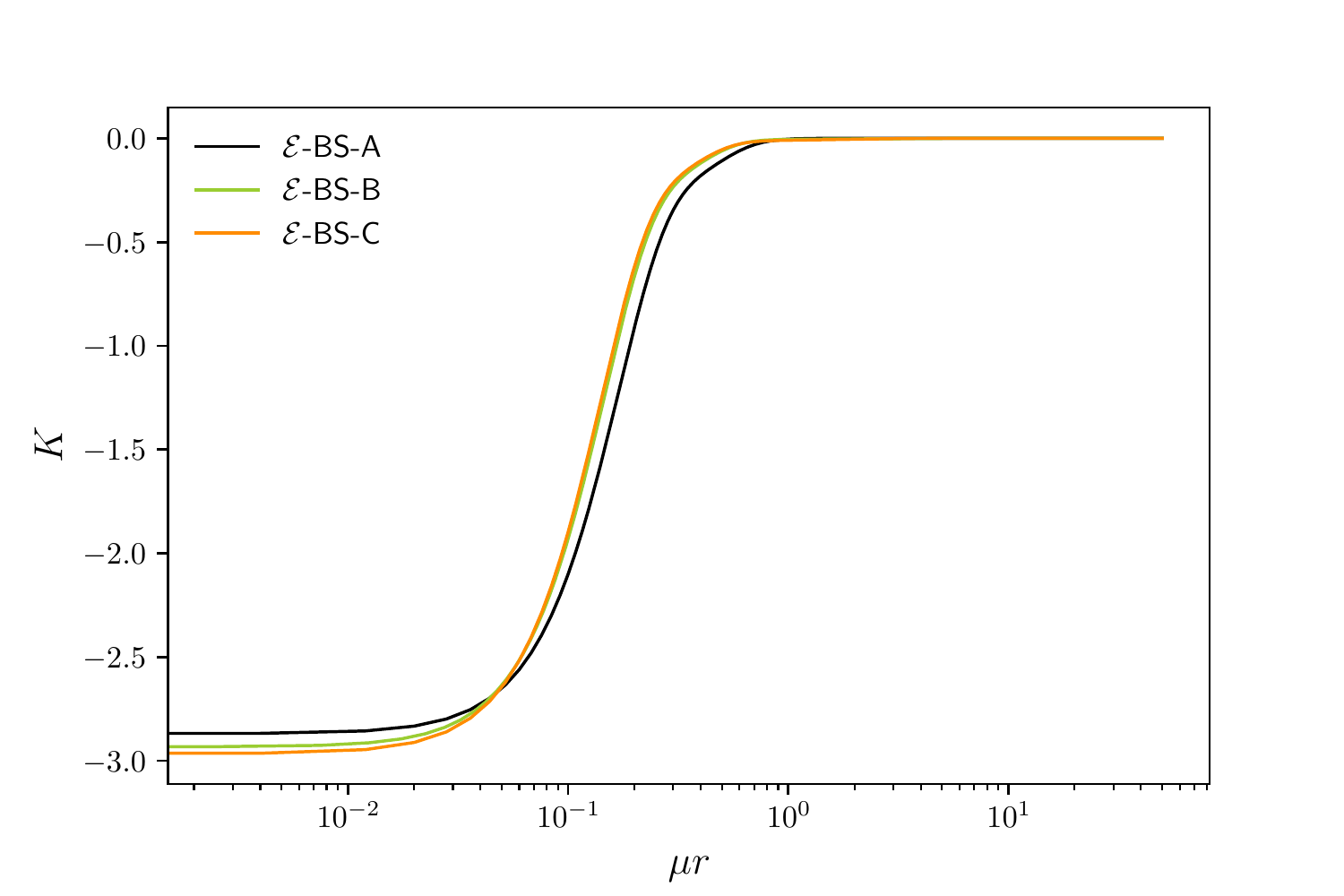}
  \includegraphics[width=0.49\textwidth]{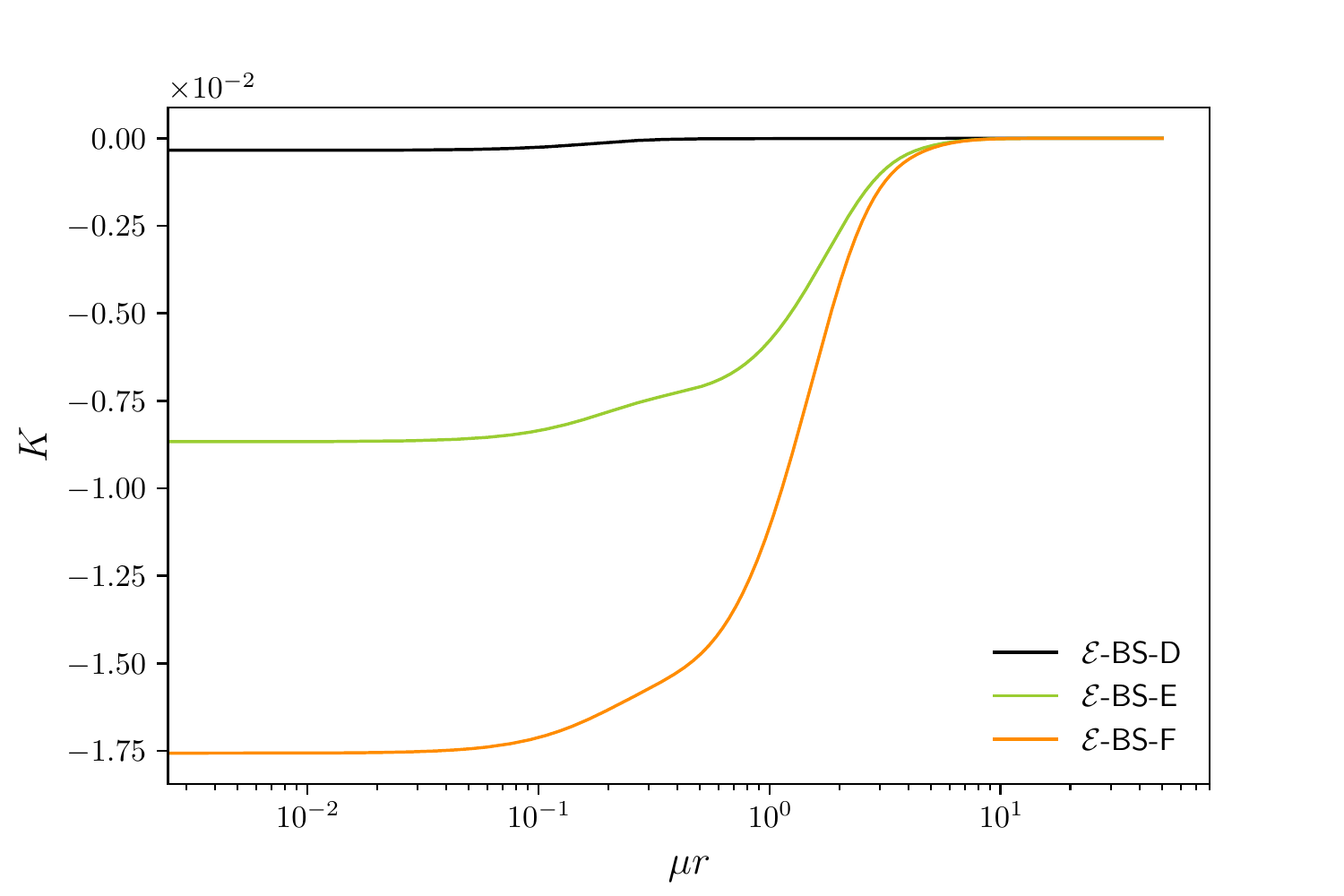}
  \caption{Kretschmann scalar for ${\cal E}$-boson star configurations  described in table \ref{tab:data}.}
\label{fig:K} 
\end{figure}

\begin{figure}
\centering
  \includegraphics[width=0.49\textwidth]{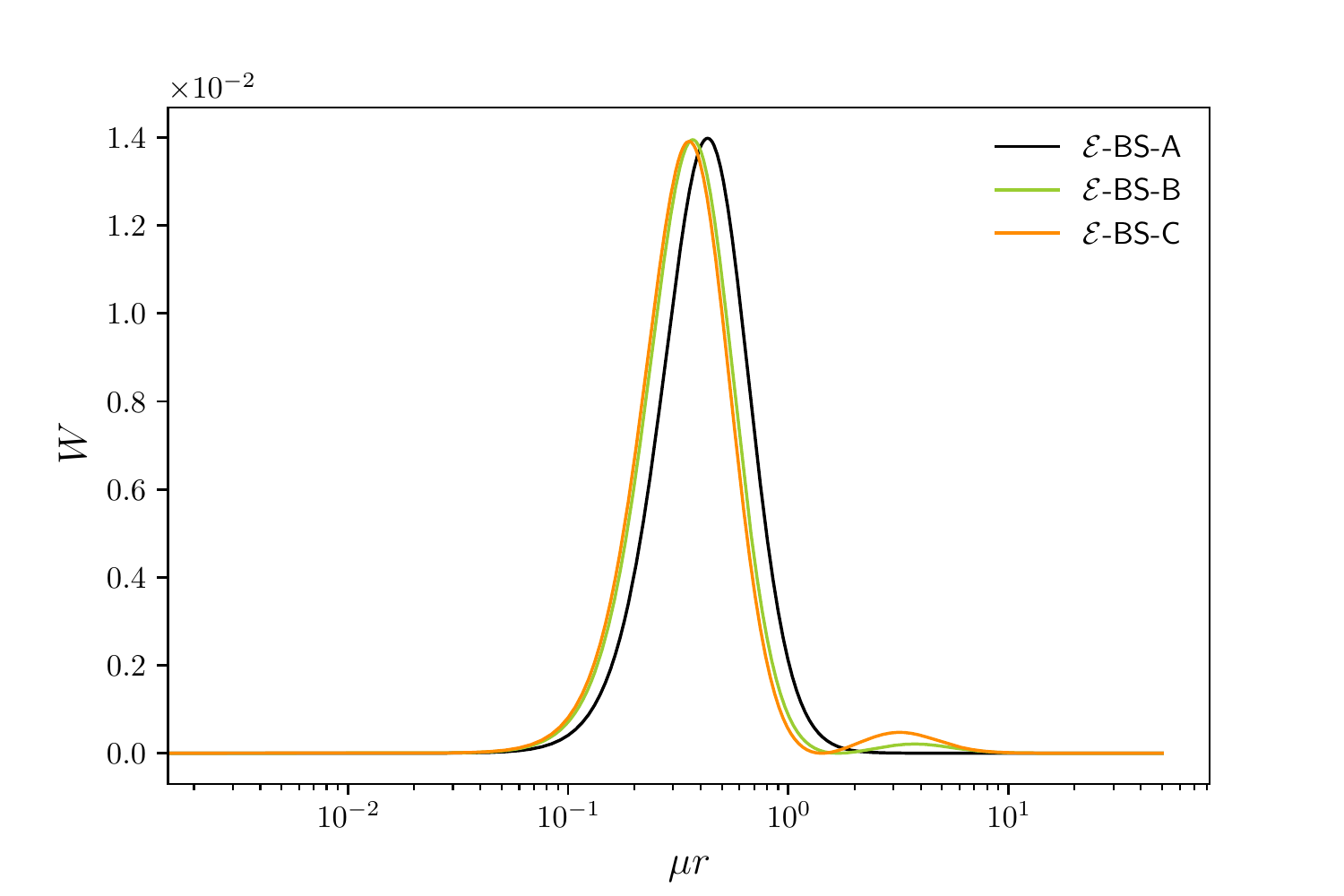}
  \includegraphics[width=0.49\textwidth]{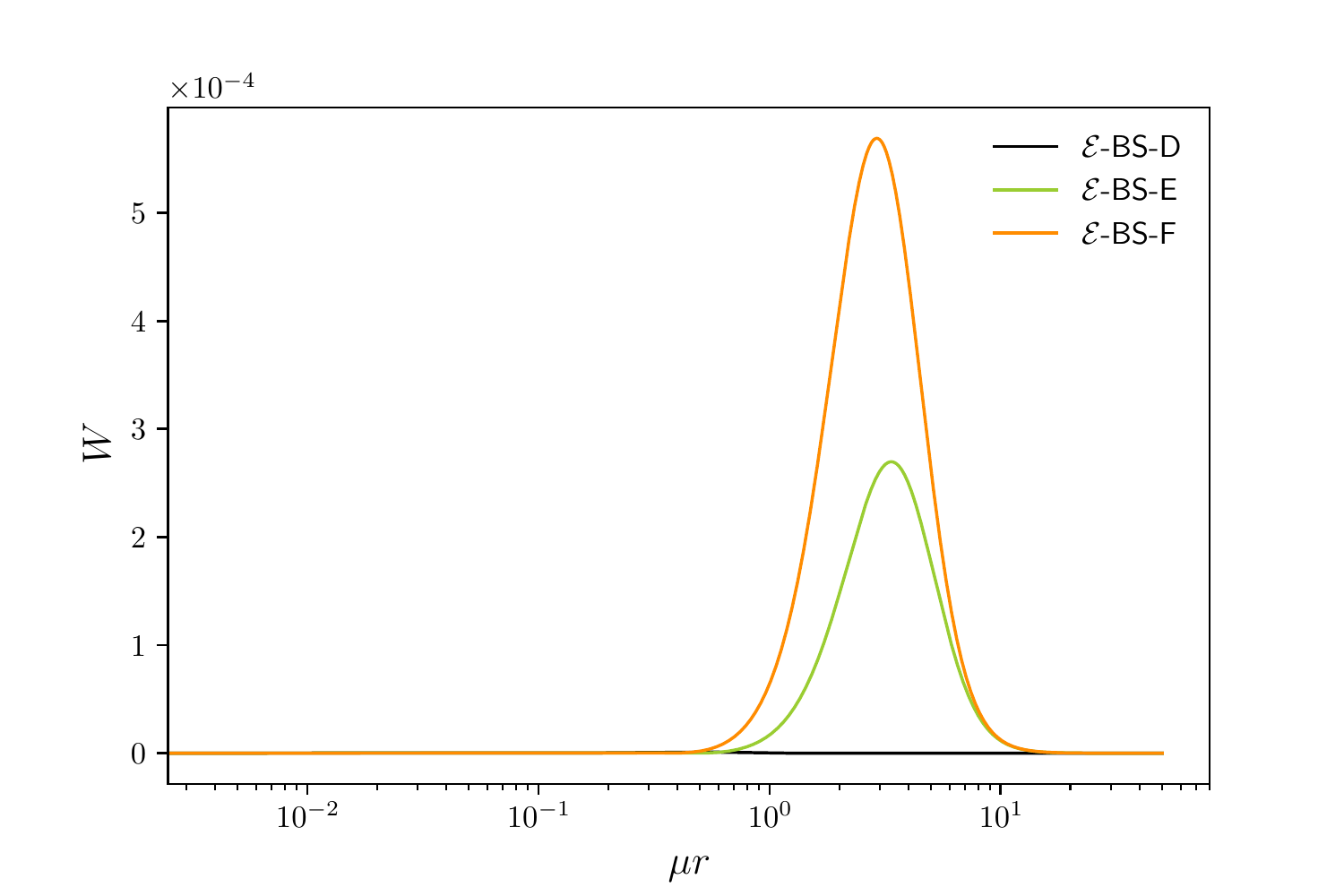}
  \caption{Weyl scalar for ${\cal E}$-boson star configurations described in table \ref{tab:data}.}
\label{fig:w} 
\end{figure}

Regarding the geometric scalars, we present their behavior for the six cases of the Table \ref{tab:data} labeled as cases ${\cal E}$-BS. Several of them describe concentric spheres and one of them (solution $\mathcal{E}$-BS-A) describe a shell-like configuration.  In all cases the scalar of curvature (which follows the total density), the Kretschmann and Weyl scalars are well-behaved Fig.~\ref{fig:Ricci}, \ref{fig:K}, \ref{fig:w}.  For cases A-C, all curvature scalars behave in a similar manner given that we have similar quantities of exotic and canonical matter, not so for the D-F cases that shown differences in their maximum (or minimum) values. Central values $\phi_0$ and $\chi_0$ are similar in magnitude, but the curvature behavior is affected by the content of canonical scalar field, smaller values for $\phi_0$ results in flatter spaces.

\begin{table}
\centering
\begin{tabular}{|c|c|c|c|c|c|c|c|}
\hline
\textit{}&\multicolumn{7}{ c |}{$\mu_\varphi=\mu_\chi=\mu$}{}\\
\hline
Solution &$\phi_0$ & $\chi_0$ & $\omega/\mu$&$\lambda$ & $\mu M$&$\mu R_{99}$&$C$\\
\hline
BS-A &$1.057\times10^{-2}$& 0 & 0.975 & - & 0.3696 & 24.27 & $1.523\times10^{-2}$\\
BS-B &$4.650\times10^{-2}$& 0 & 0.901 & - & 0.6027 & 10.68 & $5.645\times10^{-2}$\\
Ph-A & 0 & 0.1 & - & 925.8 & $-0.0102$ & 2.903 & $-3.519\times10^{-3}$\\
Ph-B & 0 & 1.0 & - & 2.559 & $-11.44$ &  6.665 & $-1.743$ \\
${\cal E}$-BS-A & $9.951\times10^{-3}$ & 0.1 & 0.975 & 925.6 & 0.3793 & 24.80 & $1.530\times10^{-2}$ \\
${\cal E}$-BS-B & $4.512\times10^{-2}$ & 0.1 & 0.901 & 925.6 & 0.6148 & 10.81 & $5.688\times10^{-2}$ \\
${\cal E}$-BS-C & $5.409\times10^{-2}$ & 0.1 & 0.885 & 927.2 & 0.6322 & 9.687 & $6.526\times10^{-2}$\\
${\cal E}$-BS-D & $1.033\times10^{-2}$ & 0.01 & 0.975 & $9.403\times 10^{4}$ & 0.3662 & 24.57 & $1.491\times10^{-2}$\\
${\cal E}$-BS-E & $4.698\times10^{-2}$ & 0.01 & 0.901& $9.411\times 10^{4}$ & 0.6040 & 10.62 & $5.689\times10^{-2}$\\
${\cal E}$-BS-F & $5.539\times10^{-2}$& 0.01 & 0.886& $9.430\times 10^{4}$ & 0.6196 & 9.614 & $6.445\times10^{-2}$\\
\hline
\end{tabular}
\caption{
}
\label{tab:data}
\end{table}

%%%%%%%%%%%%%%%%%%%%%%%
%%%   Final Remarks   %%%
%%%%%%%%%%%%%%%%%%%%%%%

\section{Final remarks}
\label{Sec:Final_remarks}

We have designed a configuration such that the exotic matter, described by a real massive scalar field with self-interaction and such that the corresponding stress energy tensor has a global sign opposite to the one corresponding to a canonical scalar field, is distributed inside a usual, canonical boson star configuration described by a massive complex scalar field.

We have been able to solve the corresponding differential equations considering that the configuration is static and has spherical symmetry, using an integration method based on the spectral solver procedure and demanding regularity at the origin and asymptotic flatness. We obtain several examples of configurations based on three free parameters, namely the central amplitudes of both scalar fields and the ratio between the exotic scalar mass $\mu_\chi$ and the canonical one $\mu_\varphi$.  

From the configurations obtained, several interesting features were deduced. First of all, the static spherical configurations inherit a part of the properties of boson stars and phantom solitons, however the gravitational (repulsive) interaction between the fields creates new global features of the configurations, such as an appreciable increase in compactness and also different morphologies, with the phantom field always within the canonical field.  
As in boson stars, it is obtained that in the plots of the total mass of the configuration versus the frequency, have the usual snail-like plot, with the possible implication  that the maximum of each plot separates a region with stable total configurations from the unstable ones.

We also found another interesting feature regarding the ratio of the central amplitudes of the exotic scalar field, $\chi_0$, to the corresponding of the canonical one, $\phi_0$; when the ratio is much smaller than one, both distributions, the exotic and the canonical one, are concentric spheres, however, as the ratio grows, the canonical distribution seems to be pushed outwards form the center, forming a shell-like distribution containing in the center the exotic field. 

We obtained the scalars of curvature which are regular at all points and describe the characteristic features of the different types of matter on the geometry. An eloquent feature of $\cal E$-boson stars is that it is possible to tune the amplitudes of the fields in such a way that the NEC is always satisfied and stable numerical implementation could be obtained even with a phantom scalar field. Further information will be given in the forthcoming part II of our work.

We have thus designed and obtained interesting configurations with two types of scalar fields. Clearly, the dynamics of such configurations, with some possible stable or stationary states, is a most pressing question which will be deal with in a forthcoming work. Preliminary results show that some of these configurations remain bounded.

%%%%%%%%%%%%%%%%%%%%%%%%%%%
%%%   ACKNOWLEDGMENTS   %%%
%%%%%%%%%%%%%%%%%%%%%%%%%%%
\section*{Acknowledgements}
This work was partially supported by 
DGAPA-UNAM through grants IN110218 and IN105920, by the CONACyT Network Project No. 376127 ``Sombras, lentes y ondas gravitatorias generadas por objetos compactos astrof\'\i sicos". VJ and EJ acknowledge financial support from CONACyT graduate grant program.

%%%%%%%%%%%%%%%%%%%%%%%%%%%%%%%%%%%%%%%%%%%%%%%%%%
\appendix
%%%%%%%%%%%%%%%%%%%%%%%%%%%%%%%%%%%%%%%%%%%%%%%%%%

%%%%%%%%%%%%%%%%%%%%%%%%%%%%%%%%%%%%%%%%%%%%%%%%%%
\section{Solutions with \texorpdfstring{$\mu_\varphi\neq\mu_\chi$}{mphi different from mchi}}
\label{Sec:diff_mu}
%%%%%%%%%%%%%%%%%%%%%%%%%%%%%%%%%%%%%%%%%%%%%%%%%%

The ratio between the mass parameter of $\chi$ and $\varphi$ is also a free parameter in the model and cannot be absorbed by any simple rescaling. Nevertheless the implication of modifying this parameter are relatively simpler than the effects of the $\chi_0$ parameter. 

\begin{figure}
  \includegraphics[width=0.49\textwidth]{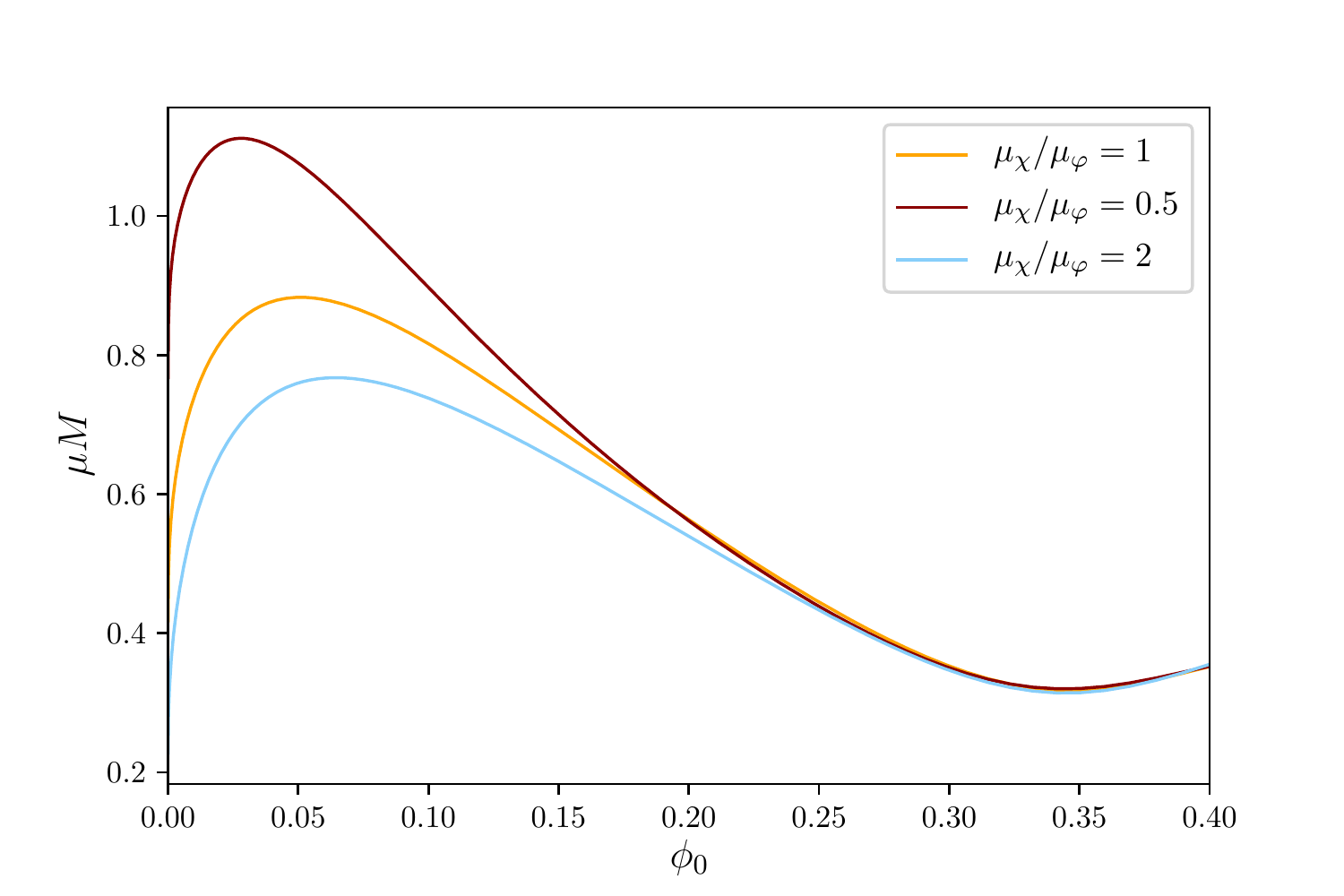} \includegraphics[width=0.49\textwidth]{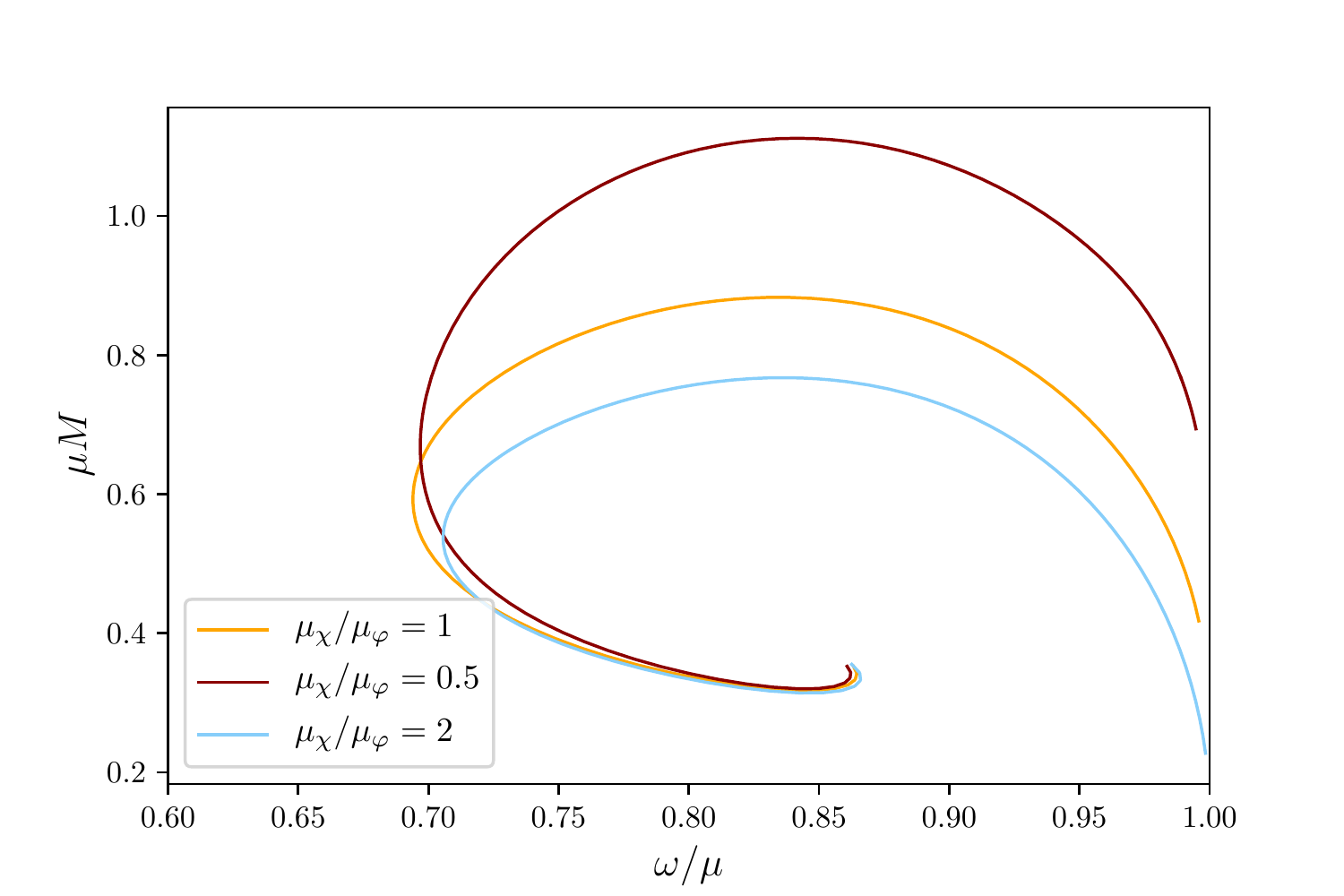}
  \caption{$\chi_0=0.4$ configurations with three different values for the ratio $\mu_\chi/\mu_\varphi$. Left panel:  Value of the canonical scalar field $\phi$ at $r=0$ \textit{vs}. total mass. Right panel: Total mass \textit{vs}. frequency.
}
\label{fig:omegavsM_muchi} 
\end{figure}

\begin{figure}
  \includegraphics[width=0.49\textwidth]{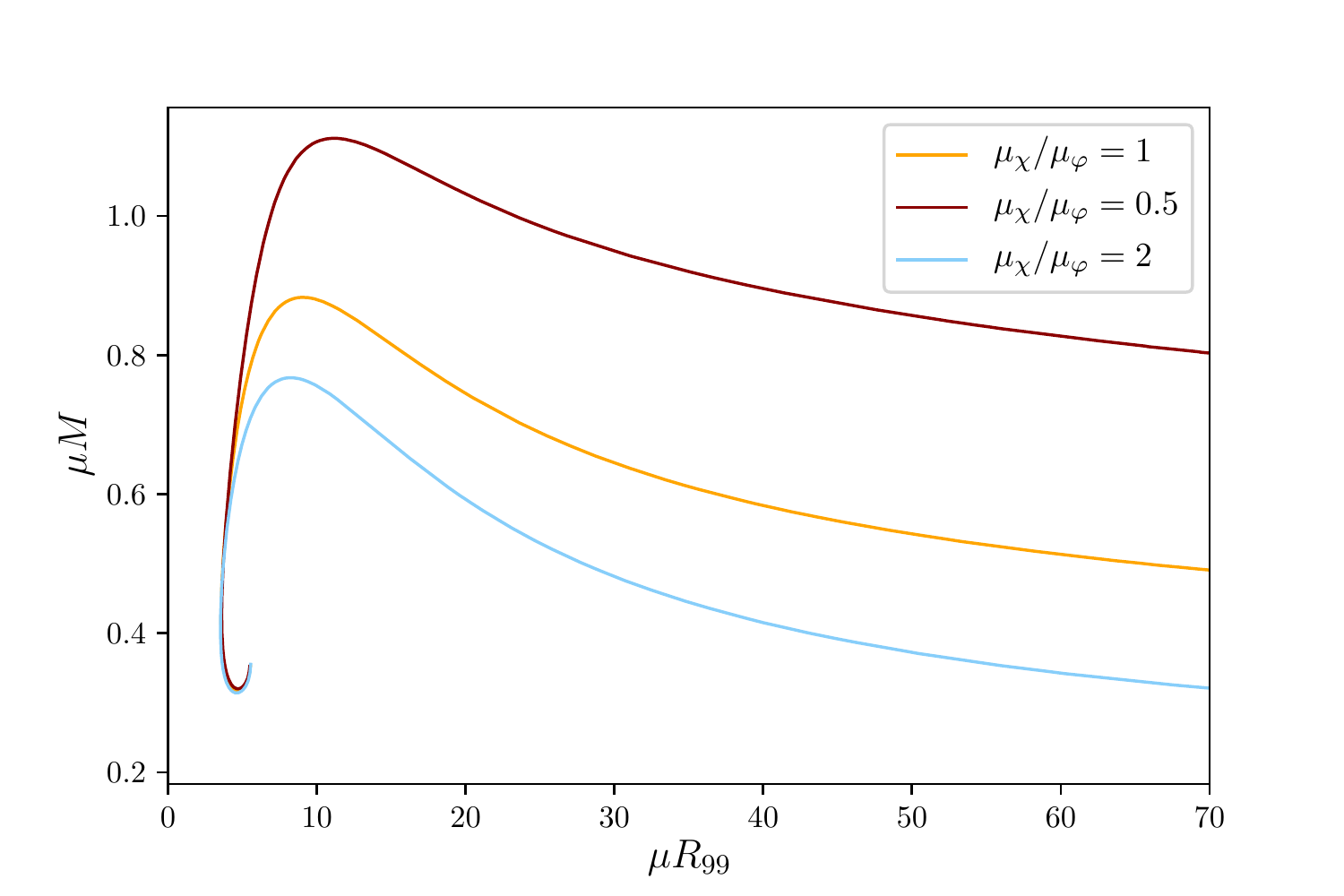} \includegraphics[width=0.49\textwidth]{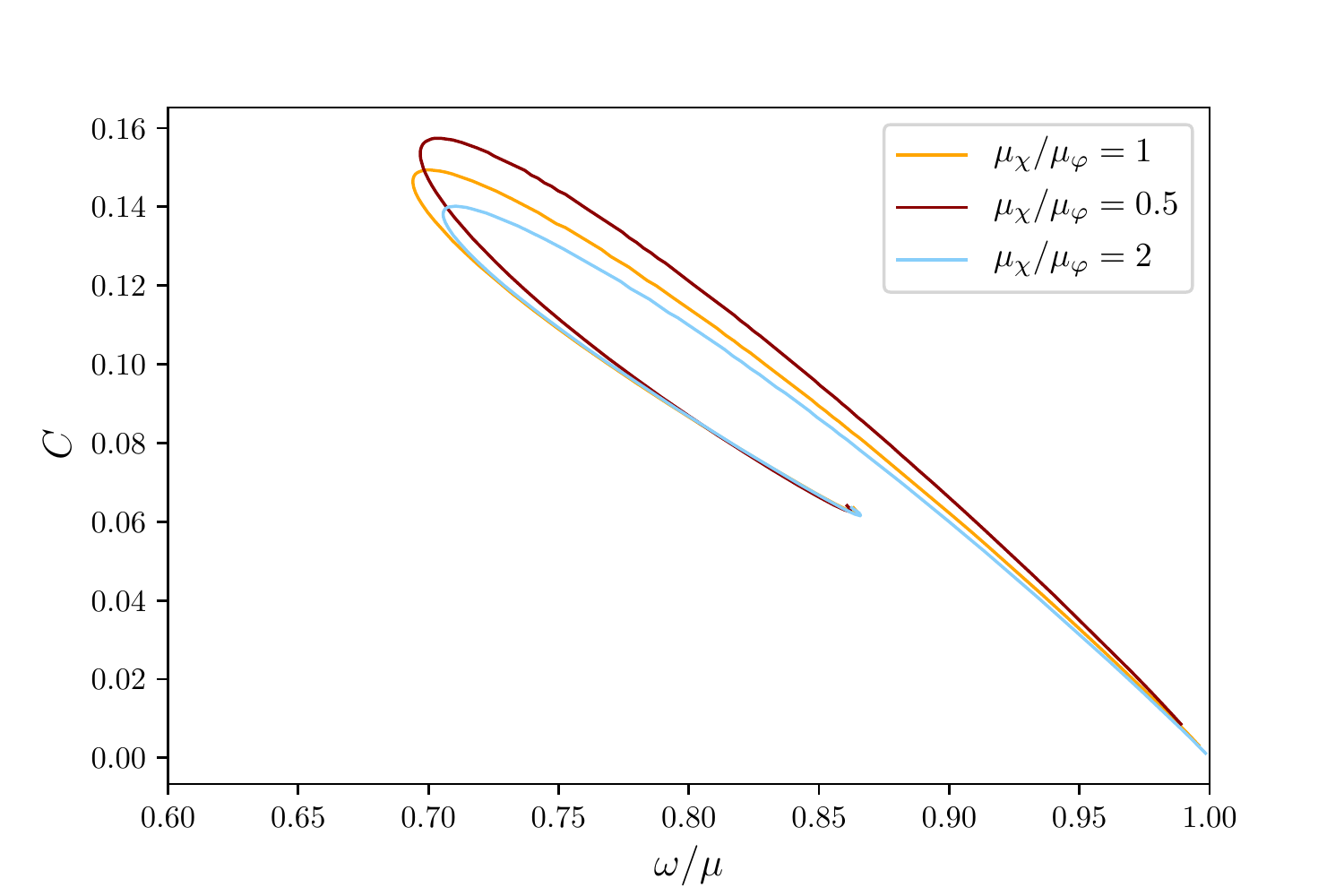}
  \caption{$\chi_0=0.4$ configurations with three different values for the ratio $\mu_\chi/\mu_\varphi$. First panel: Radius $R_{99}$ \textit{vs}. total mass. Second panel: Frequency \textit{vs}. compactness.
}
\label{fig:RvsM_muchi} 
\end{figure}

In Figs.~\ref{fig:omegavsM_muchi} and \ref{fig:RvsM_muchi} we show some global quantities of solutions for fixed value $\chi_0=0.4$ and three different values of the quotient $\mu_\chi/\mu_\varphi$. The conclusions regarding the mass and compactness are straightforward: as $\mu_\chi/\mu_\varphi$ increases, the effects of the phantom field are more present and bigger values for the total mass of the distribution and for the compactness are obtained. We have verified that similar effects are obtained for other values of $\chi_0$, in addition to those presented in the above figures. For example, we have configurations with $\mu_\chi/\mu_\varphi=1$ which do not exceed the Buchdahl limit, but it is possible to modulate the value of the quotient to obtain configurations with compactnesses as close to 0.5 as desired. 

%

%%%%%%%%%%%%%%%%%%%%%%
%%%   REFERENCES   %%%
%%%%%%%%%%%%%%%%%%%%%%

%\bibliographystyle{bibtex/prsty}
\bibliography{ref}

%%%%%%%%%%%%%%%
%%%   END   %%%
%%%%%%%%%%%%%%%

\end{document}